\documentclass[
 reprint,
 amsmath,amssymb,
 aps,
 pra,
 superscriptaddress,
]{revtex4-1}

\usepackage{graphicx}
\usepackage{bm}
\usepackage{hyperref}
\usepackage{braket}
\usepackage[utf8]{inputenc}
\usepackage{xcolor}

\usepackage{qcircuit}

\newcommand{\mr}[1]{\mathrm{#1}}
\newcommand{\td}[1]{\tilde{#1}}

\begin{document}
\title{Calculation of the Green's function on near-term quantum computers}
\author{Suguru Endo}
 \email{endo@qunasys.com}
 \affiliation{QunaSys Inc., Aqua Hakusan Building 9F, 1-13-7 Hakusan, Bunkyo, Tokyo 113-0001, Japan}
 
\author{Iori Kurata}
 \email{iori-kurata636@g.ecc.u-tokyo.ac.jp}
 \affiliation{Department of Applied Physics, The University of Tokyo, Tokyo 113-8656, Japan}
 
\author{Yuya O. Nakagawa}
 \email{nakagawa@qunasys.com}
 \affiliation{QunaSys Inc., Aqua Hakusan Building 9F, 1-13-7 Hakusan, Bunkyo, Tokyo 113-0001, Japan}
 
\date{\today}

\begin{abstract}
The Green's function plays a crucial role when studying the nature of quantum many-body systems, especially strongly-correlated systems.
Although the development of quantum computers in the near future may enable us to compute energy spectra of classically-intractable systems, methods to simulate the Green's function with near-term quantum algorithms have not been proposed yet.
Here, we propose two methods to calculate the Green's function of a given Hamiltonian on near-term quantum computers.
The first one makes use of a variational dynamics simulation of quantum systems and computes the dynamics of the Green's function in real time directly.
The second one utilizes the Lehmann representation of the Green's function and a method which calculates excited states of the Hamiltonian.
Both methods require shallow quantum circuits and are  compatible with near-term quantum computers. 
We numerically simulated the Green's function of the Fermi-Hubbard model and demonstrated the validity of our proposals.
\end{abstract}

\maketitle

\section{Introduction}
The advent of a primitive but still powerful form of quantum computers, called noisy intermediate-scale quantum (NISQ) devices, is approaching~\cite{Preskill2018}.
NISQ devices have a few hundreds to thousands of qubits under highly precise control but they are not fault-tolerant.
It is believed that the behaviour of NISQ devices will soon reach a stage that one cannot simulate its dynamics using classical computers, due to exponentially-increasing size of the Hilbert space with the number of qubits~\cite{harrow_2017, boixo_2018, villalonga_2019, chen_201802, chen_201805}.
Therefore, many researchers expect that NISQ devices will exhibit supremacy over classical computers for some specific tasks, even though they cannot execute complicated quantum algorithms requiring a huge number of qubits and gate operations due to the erroneous nature of them~\cite{Preskill2018}.

One of the most promising tasks in which near-term quantum computers may outperform classical computers is quantum simulation, where one computes energy eigenvalues and/or eigenstates of a given quantum system.
It enables us to calculate and predict properties of quantum many-body systems, which is of great importance to many fields such as quantum chemistry, condensed matter physics, and material science~\cite{mcardle2018quantum, Cao2018}.
The most celebrated algorithm for quantum simulation on near-term quantum computers is the variational quantum eigensolver (VQE)~\cite{peruzzo2014variational,kandala2017hardware,moll2018quantum,mcclean2016theory}, in which energy eigenstates and eigenenergies of the system are obtained based on the variational principle of quantum mechanics.
Although the VQE was originally proposed for finding only the ground state, its extension to excited states was discussed in several recent works~\cite{nakanishi2018subspace,higgott2019variational,jones2019variational,parrish2019quantum,PhysRevA.95.042308}.

However, other important quantities to investigate quantum many-body systems other than eigenenergies and eigenstates have remained relatively disregarded in the recent development of near-term quantum algorithms, i.e., the Green's function and the spectral function~\cite{bonch2015green,abrikosov2012methods,fetter2012quantum}.
They are fundamental to study quantum many-body systems, especially strongly-correlated systems;
for example, in condensed matter physics, the spectral function tells us that the dispersion relation of quasipaticle excitations of a system, which gives crucial information on  high-$T_c$ superconductivity~\cite{ding1996spectroscopic}, magnetic materials~\cite{coey2010magnetism}, and topological insulators~\cite{hasan2010colloquium}. 
While several methods based on the Suzuki-Trotter decomposition of the time evolution operator~\cite{Bauer2016,kreula2016non} or quantum phase estimation~\cite{wecker2015solving,Roggero2019, kosugi2019construction} were already proposed to calculate the Green's function on quantum computers (including general multi-point time correlation functions~\cite{Pedernales2014} used in gauge theories and nuclear physics~\cite{Lamm2019, Mueller2019, Lamm2019Parton}),  they require a large number of qubits and gate operations which are hard to realize with near-term quantum computers.

In this paper, we propose two different algorithms for evaluating the Green's function on near-term quantum computers.
The first one takes advantage of the variational quantum simulation (VQS) algorithm~\cite{li2017efficient,yuan2018theory,mcardle2019variational,endo2018variational,heya2019subspace} for an efficient calculation of the Green's function in real time.
We extend the original VQS algorithm in order to calculate the transition amplitude of general quantum operators between two different quantum states after time evolution.
The second one is based on the Lehmann representation of the spectral function~\cite{bonch2015green,abrikosov2012methods,fetter2012quantum}.
By calculating excited states of a given system and evaluating the transition amplitude of appropriate operators by using the subspace-search variational quantum eigensolver (SSVQE)~\cite{nakanishi2018subspace} or the multistate contracted VQE (MCVQE)~\cite{parrish2019quantum}, one can compute the spectral function (hence the Green's function).
We confirm the validity of our methods using numerical simulations of the Fermi-Hubbard model, a model of strongly-correlated system.
The extension of our methods to finite temperature and general correlation functions is also discussed.

The rest of the paper is organized as follows.
In Sec.~\ref{sec:Review}, we briefly review the definition of the Green's function and the spectral function at zero temperature.
In Sec.~\ref{sec:VQS}, we propose a method to calculate them by using the VQS algorithm.
We describe another method to calculate the Green's function and the spectral function as a simple application of the SSVQE algorithm~\cite{nakanishi2018subspace} and the MCVQE algorithm~\cite{parrish2019quantum} in Sec.~\ref{sec:SSVQE}.
Sec.~\ref{sec:Numerics} is dedicated to demonstration of our methods by performing numerical simulations calculating the spectral functions of the 2-site Fermi-Hubbard model. Sec.~\ref{app:dependence_n_d} discusses how the depth of the ansatz for the VQS affects the accuracy of numerical simulations.
In Sec.~\ref{sec:gatecount}, we discuss the feasibility of implementing our proposed algorithms on near-term quantum computers from the viewpoint of the number of gate operations and their error rate required. 
We present the extension of our methods to the finite temperature Green's function in Sec.~\ref{sec:FiniteT}.
We discuss implications and possible future directions of our results, and make a conclusion in Sec.~\ref{sec:Conclusion}.
Appendix~\ref{appendix:resource} provides detailed resource estimations of our algorithms. In Appendix~\ref{appendix:gatecount}, we describe details of the discussion in Sec.~\ref{sec:gatecount}.  Appendix~\ref{additional} provides additional numerical results, e.g., four-site Hubbard model simulation.
Appendix~\ref{App: details} gives details of the numerical simulations.

\section{Review of the Green's function and the spectral function at zero temperature \label{sec:Review}}
In this section, we briefly review definition of the Green's function at zero temperature for consistency~\cite{bonch2015green,abrikosov2012methods,fetter2012quantum}.

Let us consider a fermionic system described by Hamiltonian $H$ which is composed of fermionic creation (annihilation) operators $c_a, c_a^\dag$, where $a$ is a label to specify the fermionic mode.
For example, $a$ can be $(k,\sigma)$, where $k$ denotes the momentum and $\sigma=\uparrow, \downarrow$ denotes the spin of the fermion.
The retarded Green's function at zero temperature is defined as
\begin{align}
  G^R_{ab}(t)  = -i \Theta(t) \braket{c_a(t) c_b^{\dagger}(0) + c_b^{\dagger}(0)c_a(t) }_0,
\end{align}
where $\Theta(t)$ is the Heaviside step function, $c_a(t) = e^{iHt} c_a e^{-iHt}$ is the Heisenberg representation of the operator $c_a$, and $\braket{\ldots}_0 = \braket{\mr{G}|\ldots|\mr{G}}$ denotes the expectation value by the ground state of the Hamiltonian, $\ket{\mr{G}}$.
We employ the natural unit where the Plank constant $\hbar$ and the Boltzmann constant $k_B$ are $\hbar=k_B=1$.
For simplicity, throughout this paper we consider the Green's function $G^{R}_{ab}$ with $a=b=(k, \uparrow)$, namely, the Green's function in the momentum space with identical spin.
We simply write
\begin{align}
  G^R_{(k,\uparrow),(k,\uparrow)}(t)  = G^R_k(t)
\end{align}
in all other parts of the paper.
We note that extension of proposed methods in this study to the Green's function with general indices is straightforward.

The Green's function is related to another important physical quantity to investigate quantum many-body systems, namely, the spectral function $A_k(\omega)$.
It is defined through the Fourier transform of $G^R_k(t)$,
\begin{eqnarray}
\td{G}^R_k(\omega)  = \int_{-\infty}^{\infty} dt \, e^{i(\omega+i\eta)t} G^R_k(t) \\
=: \int_{-\infty}^{\infty} d\omega' \frac{A_k(\omega')}{\omega - \omega' + i\eta},
\end{eqnarray}
where $\eta \to +0$ is a factor to assure the convergence of the integral.
The spectral function and the Green's function $\td{G}_k^R(\omega)$ have a relation
\begin{align}
 A_k(\omega) = - \pi^{-1} \mr{Im} \, \td{G}_k^R(\omega).
\end{align}
Finally, we introduce the Lehmann representation of the spectral function which utilizes the energy eigenvalue $E_n$ and the eigenstate of the Hamiltonian $\ket{E_n}$,
\begin{align}
 \begin{aligned}
  A_k(\omega) &=& \sum_{n} \left( \frac{ |\braket{E_n|c_k^\dag|\mr{G}}|^2 }{\omega+E_\mr{G}-E_n+i\eta}  \right. \\
  && \left. + \frac{ |\braket{E_n|c_k|\mr{G}}|^2 }{\omega-E_\mr{G}+E_n+i\eta} \right),
 \end{aligned}
 \label{Eq:Lehmann}
\end{align}
where we call the first (second) term as particle (hole) part of the spectrum function.

In Sec.~\ref{sec:VQS}, we compute the spectral function by performing the Fourier transformation to $G_k^R(t)$ calculated by the VQS-based method while we compute it by the Lehmann representation~\eqref{Eq:Lehmann} with quantities calculated by the VQE-based method in Sec.~\ref{sec:SSVQE}.

\section{Computation of Green's function with variational quantum simulation \label{sec:VQS}}
In this section, we first review the VQS algorithm~\cite{li2017efficient,yuan2018theory,mcardle2019variational,endo2018variational} which calculates a quantum state after the time evolution by a given Hamiltonian within parametrized ansatz states.
Next we propose the method to compute the Green's function by extending the original VQS algorithm.

\subsection{Review of variational quantum simulation}
Here we review the variational quantum real time simulation algorithm introduced in Ref.~\cite{li2017efficient}. Let us consider an ansatz quantum state created by a parametrized quantum circuit,
\begin{align}
\ket{\psi (\vec{\theta})} = U(\vec{\theta}) \ket{\varphi_0} = U_{N_\theta}(\theta_{N_\theta})\dots U_{i}(\theta_i)\dots U_{1}(\theta_1)\ket{\varphi_0},
\label{eq:ansatz}
\end{align}
where $U_i(\theta_i)$ is some unitary gate with (real-valued) parameter $\theta_i$, $N_\theta$ is the number of parameters, and $\ket{\varphi_0}$ is a reference state to create the ansatz state.
We assume that $U_i(\theta_i)$ is composed of a set of Pauli rotation gates, $e^{i\alpha^{(j)}\theta_i \mathcal{P}^{(j)}}$ with a coefficient $\alpha^{(j)}$ and a Pauli matrix $\mathcal{P}^{(j)}$, and other non-parametrized gates.
For a given initial state $\ket{\psi(\vec{\theta}(0))}$ and Hamiltonian $H$, the VQS algorithm finds the solution of the Schr\"odinger equation, 
\begin{align}
 \frac{d \ket{\psi(\vec{\theta}(t))}}{dt}=-i H \ket{\psi(\vec{\theta}(t))},
 \label{schrodinger}
\end{align}
within the Hilbert space spanned by the ansatz quantum state, $\{ \ket{\psi(\vec{\theta})} \}_{\vec{\theta}}$.
Specifically, the time evolution described by Eq.~\eqref{schrodinger} is mapped to the time evolution of parameters $\theta(t)$. 
Although there are several variational principles to map Eq.~\eqref{schrodinger} to the equations for $\theta(t)$~\cite{frenkel1935wave,dirac_1930,mclachlan1964variational}, we choose McLachlan's variational principle~\cite{mclachlan1964variational} in this paper because it is the most stable and physically reasonable among them~\cite{yuan2018theory}.

McLachlan's variational principle ~\cite{mclachlan1964variational} maps Eq.~(\ref{schrodinger}) to the equation of motion of the parameters $\vec{\theta}(t)$ by minimizing the distance between the exact evolution of the Schr\"odinger equation and the evolution of the parametrized ansatz state under infinitesimal variation of time $\delta t$~\cite{li2017efficient}, 
\begin{align}
 \min \, \delta \left\lVert \left( \frac{\partial}{\partial t} + iH \right) \ket{\psi (\vec{\theta}(t))} \right\rVert, 
\end{align}
where $\|\ket{\varphi} \|=\braket{\varphi|\varphi}$ is the norm of $\ket{\varphi}$.
One can explicitly write down the equation determining $\{\dot{\theta}_i\}_i$,
\begin{align}
 \sum_j M_{i,j}\dot{\theta}_j=V_i, 
 \label{Eq:MthetaV}
\end{align}
where
\begin{equation}
 \begin{aligned}
	  M_{i,j} &=\mathrm{Re} \bigg(\frac{\partial \bra{\psi(\vec{\theta}(t))}}{\partial \theta_i}\frac{\partial \ket{\psi(\vec{\theta}(t))}}{\partial \theta_j}\bigg),\\
      V_i &=\mathrm{Im}\bigg(\bra{\psi(\vec{\theta}(t))}H \frac{\partial \ket{\psi(\vec{\theta}(t))}}{\partial \theta_i} \bigg),
 \label{Eq:MV}
 \end{aligned}
\end{equation}
for $i=1,\ldots,N_\theta$.
We note that the matrix $M$ and vector $V$ can be efficiently obtained by measurements of quantum circuits as 
described in Refs.~\cite{li2017efficient, PhysRevResearch.1.013006}.

Finally, from the solution of Eq.~\eqref{Eq:MthetaV}, one can evaluate the parameter at time $t+ \delta t$ as $\vec{\theta}(t+\delta t) \approx \vec{\theta}(t)+ \dot{\vec{\theta}}(t) \delta t$.
By iterating this procedure from the initial parameters $\vec{\theta}(0)$, we can obtain the time evolution of the parameters $\vec{\theta}(t)$ and the quantum state $\ket{\psi (\vec{\theta}(t))}$.
We note that the VQS algorithm can be viewed as approximating the time evolution operator $U(t) = e^{-iHt}$ by the variational quantum circuit $U(\vec{\theta}(t))$, although the approximation is valid only when the operators are applied for the initial state $\ket{\psi(\vec{\theta}(0))}$: $U(\vec{\theta}(t))$ is optimised to the chosen initial state $\ket{\psi(\vec{\theta}(0))}$ and $U(\vec{\theta}(t))\ket{\psi_1} \neq e^{-iHt}\ket{\psi_1}$ for a different initial state $\ket{\psi_1}$ in general.

\subsection{Computation of Green's function}
Now, we discuss how we can apply the VQS algorithm for evaluating the Green's function.
What we want to evaluate is the retarded Green's function for $t>0$,
\begin{equation}
 \begin{aligned}
 G_k^R (t) & = -i \left( \bra{\mathrm{G}} e^{i H t} c_{k \uparrow} e^{-i H t} c_{k \uparrow}^\dag  \ket{\mathrm{G}} \right. \\
& \left. + \bra{\mathrm{G}}c_{k \uparrow}^\dag e^{i H t} c_{k \uparrow} e^{-i H t} \ket{\mathrm{G}} \right).
 \label{Eq:Gfrealtime}
 \end{aligned}
\end{equation}
The first and second term can be evaluated similarly, so we focus on the first term.

First, we prepare the (approximate) ground state of a given Hamiltonian $H$ described by $N$ qubits on near-term quantum computers, by using the conventional VQE method~\cite{peruzzo2014variational,kandala2017hardware,moll2018quantum,mcclean2016theory} or the variational imaginary time simulation algorithm~\cite{mcardle2019variational,stokes2019quantum}.
We denote it as $\ket{\mr{G}} = U_G \ket{\varphi_0}$ with a unitary $U_G$ and a reference state $\ket{\varphi_0}$.
Next, we decompose the fermion operator $c_{k,\uparrow}$ into a sum of Pauli matrices~\cite{seeley2012bravyi,bravyi2002fermionic,mcardle2018quantum, Cao2018, Jordan1928},
\begin{align}
 c_{k,\uparrow} \to \sum_{n=1}^{N_k} \lambda^{(k)}_n P_n, \:\:
 c_{k,\uparrow}^\dag \to \sum_{n=1}^{N_k} \lambda^{(k)*}_n P_n,
 \label{eq:ExpandElectron}
\end{align}
where $P_n$ is a tensor product of Pauli matrices acting on $N$ qubits that satisfies $P_n^\dag=P_n, \: P_n^2=I_N$ and $\lambda^{(k)}_n$ is a (complex) coefficient.
For example, one can adopt the Jordan-Wigner transformation~\cite{Jordan1928} for the decomposition.
The first term of Eq.~(\ref{Eq:Gfrealtime}) can be rewritten as 
\begin{align}
\sum_{i, j} \lambda^{(k)}_i \lambda^{(k)*}_j \bra{\mr{G}} e^{i H t} P_i e^{- i H t} P_j \ket{\mr{G}}.
\label{Eq:decomposedGF}
\end{align}
Therefore, the problem reduces to finding a way to compute $\bra{\mr{G}} e^{i H t} P_i e^{- i H t} P_j \ket{\mr{G}}$ on near-term quantum computer.

\subsubsection{Direct method to compute Green's function}
\begin{figure}
\centering
\includegraphics[width=.45\textwidth]{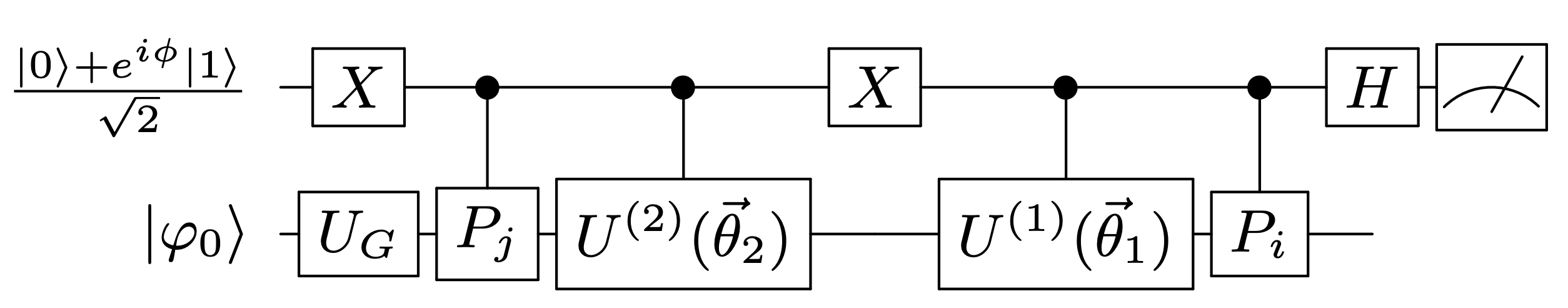}
\caption{Quantum circuit to compute Eq.~(\ref{Eq:greenterm}).
The upper horizontal line represents the ancillary qubit and the lower line does the qubits for the system of interest.
The initial state for the ancillary qubit is taken as $2^{-1/2}(\ket{0}+e^{i\phi}\ket{1})$.
The expectation value of $Z$-measurement on the ancillary qubit yields $\mr{Re}\left( e^{i\phi}\bra{\mr{G}}U^{(1)}(\vec{\theta}_1)^\dag P_i U^{(2)}(\vec{\theta}_2) P_j \ket{\mr{G}} \right)/2$.
Hence by choosing $\phi=0, \pi/2$, we can measure both the real and imaginary part of Eq.~(\ref{Eq:greenterm}).} 
\label{fig:naive circuit}
\end{figure}

A direct way to evaluate $\bra{\mr{G}} e^{iHt} P_i e^{- iHt} P_j \ket{\mr{G}}$ by the VQS algorithm is as follows.
The time evolution of the states $\ket{\mr{G}}$ and $P_j \ket{\mr{G}}$ are approximated on quantum computers as
\begin{align}
 e^{-iHt}  \ket{\mr{G}} \approx U^{(1)}(\vec{\theta}_1(t))  \ket{\mr{G}},  \\
 e^{-iHt} P_j \ket{\mr{G}} \approx U^{(2)}(\vec{\theta}_2(t)) P_j \ket{\mr{G}},
\end{align}
where $U^{(1,2)}(\vec{\theta})$ are parametrized unitary circuits and the initial parameters $\vec{\theta}_1(0)$ and $\vec{\theta}_2(0)$ of the VQS are taken so as to satisfy  $U^{(1,2)}(\vec{\theta}_{1,2}(0))=I_N$. Note that, as $U^{(1)}(\vec{\theta}_1(t))$ and $U^{(2)}(\vec{\theta}_2(t))$ are optimised for initial states $\ket{\mathrm{G}}$ and $P_j \ket{\mathrm{G}}$, respectively, generally $U^{(1)}(\vec{\theta}_1(t))\neq U^{(2)}(\vec{\theta}_2(t))$ holds.
Then, $\bra{\mr{G}} e^{i H t} P_i e^{- i H t} P_j \ket{\mr{G}}$ can be evaluated as the transition amplitude of $P_i$ between $e^{-iHt}\ket{\mr{G}}$ and $e^{-iHt} P_j \ket{\mr{G}}$,
\begin{eqnarray}
 && \bra{\mr{G}} e^{i H t} P_i e^{- i H t} P_j \ket{\mr{G}} \nonumber \\
 &\approx& \bra{\mr{G}} \left( U^{(1)}(\vec{\theta}_1(t)) \right)^\dag P_i U^{(2)} (\vec{\theta}_2(t)) P_j \ket{\mr{G}}. 
 \label{Eq:greenterm}
\end{eqnarray}
This can be evaluated by using the quantum circuit shown in Fig.~\ref{fig:naive circuit}. 

However, this quantum circuit necessitates a huge number of control operations from an ancillary qubit, because two controlled-unitary operations for $U^{(1)}(\vec{\theta}_1(t))$ and  $U^{(2)}(\vec{\theta}_2(t))$ are present.
Since control operations in near-term quantum computers are costly and have relatively low fidelity in general, the large number of them in the algorithm will deteriorate the performance of computations in real near-term quantum computers.
Therefore, we propose another more efficient method to evaluate $\bra{\mr{G}} e^{iHt} P_i e^{- iHt} P_j \ket{\mr{G}}$ which will safely run on near-term quantum computers.

\subsubsection{Efficient method to compute Green's function}
\begin{figure}
\centering
\includegraphics[width=.35\textwidth]{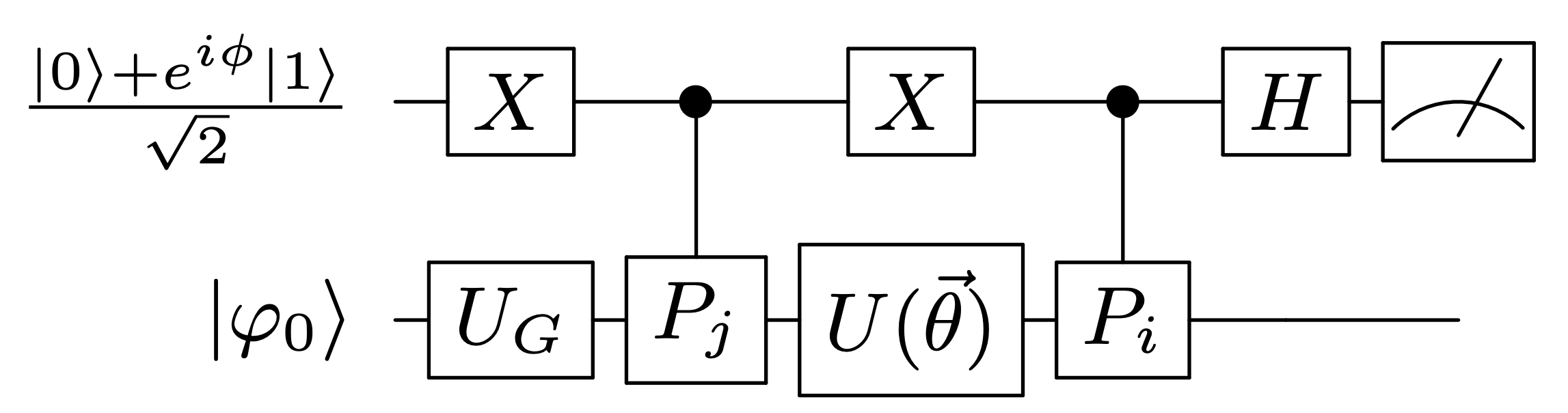}
\caption{Quantum circuit to compute Eq.~\eqref{Eq:efficient rep}.
Again, the upper horizontal line represents the ancillary qubit and the lower line does the qubits for the system of interest.
The initial state for the ancillary qubit is taken as $2^{-1/2}(\ket{0}+e^{i\phi}\ket{1})$.
The expectation value of $Z$-measurement on the ancillary qubit yields $\mr{Re}\left( e^{i\phi}\bra{\mr{G}}U(\vec{\theta})^\dag P_i U(\vec{\theta}) P_j \ket{\mr{G}} \right)/2$,
so we can measure both the real and imaginary part of Eq.~(\ref{Eq:efficient rep}) by choosing $\phi=0, \pi/2$.
The number of required controlled operations is only two in this case.
We can further eliminate controlled operations by using the method proposed in Ref.~\cite{PhysRevResearch.1.013006}.} 
\label{fig:efficient circuit}
\end{figure}

The problem in the previous method stems from the fact that the variational representations of the time evolution operator $U(t)=e^{-iHt}$ are different between two initial states, $\ket{\mr{G}}$ and $P_j\ket{\mr{G}}$.
If we can construct the variational circuit $U(\vec{\theta}(t))$ which simultaneously approximates the time evolution operator for the initial states $\ket{\mr{G}}$ and $P_j \ket{\mr{G}}$, we obtain
\begin{eqnarray}
 && \bra{\mr{G}} e^{i H t} P_i e^{- i H t} P_j \ket{\mr{G}} \nonumber \\
 &\approx& \bra{\mr{G}} \left( U(\vec{\theta}(t)) \right)^\dag P_i U(\vec{\theta}(t)) P_j \ket{\mr{G}}. 
 \label{Eq:efficient rep}
\end{eqnarray}
In this case, the quantum circuit for evaluating the term is significantly simplified as depicted in Fig.~\ref{fig:efficient circuit}.
We will now describe how to construct such variational quantum circuit which simultaneously approximates the time evolution operator for general multiple states. 

Here, we consider the most general case where we have $L$ multiple initial states for the time evolution, $\{\ket{\psi_l} \}_{l=0}^{L-1} $.
Let us consider a state with ancilla, 
\begin{align}
 \ket{\Psi_0}= \frac{1}{\sqrt{L}}\sum_{l=0}^{L-1} \ket{l}_a \otimes \ket{\psi_l}_s,
 \label{eq:ansatz for GF}
\end{align}
where subscripts $a$ and $s$ denote the ancilla and the system of interest, respectively, and $\{\ket{l}_a\}_l$ is orthonormal state of the ancilla.
We note that when $2^{k-1} < L \leq 2^k$, $k$ ancilla qubits are needed.
By using the VQS algorithm to the state $\ket{\Psi_0}$ with the variational quantum circuit $I_a \otimes U_s (\vec{\theta})$, we can construct the unitary operator $U_s(\vec{\theta}(t))$ which approximately behaves as the time evolution operator for $\{\ket{\psi_l}\}_{l=0}^{L-1}$.
To be more concrete, the ansatz state for the VQS algorithm is,
\begin{align}
\ket{\Psi(\vec{\theta})} &:= I_a \otimes U_s  (\vec{\theta}) \ket{\Psi_0} \nonumber \\
&= \frac{1}{\sqrt{L}}\sum_{l=0}^{L-1} \ket{l}_a \otimes \ket{\psi_l(\vec{\theta})}_s,
\label{eq:Ansatz for geneal VQS}
\end{align}
where we define $\ket{\psi_l(\vec{\theta})}_s= U_s (\vec{\theta})\ket{\psi_l}_s$.
The $M$ matrix and $V$ vector in Eq.~\eqref{Eq:MV} become 
\begin{align}
M_{i,j} &= \mr{Re}\bigg( \frac{ \partial \bra{\Psi(\vec{\theta}(t))}}{\partial \theta_i} \frac{ \partial \ket{\Psi(\vec{\theta}(t))}}{\partial \theta_j} \bigg) \nonumber \\
&=\frac{1}{L}\sum_{l=0}^{L-1}\mr{Re}\bigg(\frac{ \partial \bra{\psi_l(\vec{\theta}(t))}_s}{\partial \theta_i} \frac{\partial \ket{ \psi_l(\vec{\theta}(t))}_s}{\partial \theta_j} \bigg),
\label{eq:M_with_ancilla}
\end{align}
and 
\begin{align}
V_i &= \mr{Im}\bigg(\bra{\Psi(\vec{\theta}(t))} H \frac{ \partial \ket{\Psi(\vec{\theta}(t))}}{\partial \theta_i} \bigg) \nonumber \\
&=\frac{1}{L}\sum_{l=0}^{L-1} \mr{Im} \bigg(\bra{\psi_l(\vec{\theta}(t))}_s H \frac{ \partial \ket{\psi_l(\vec{\theta}(t))}_s}{\partial \theta_i} \bigg).
\label{eq:V_with_ancilla}
\end{align}
From this expression, one notice that this algorithm minimizes the average of $\delta \left\lVert \left( \partial/ \partial t + iH \right) \ket{\psi_l (\vec{\theta}(t))} \right\rVert$ for $l=0,\ldots,L-1$.
We also note that the algorithm itself can run without resorting to the ancilla because each summand in Eqs.~\eqref{eq:M_with_ancilla}, \eqref{eq:V_with_ancilla} can be computed by a distinct quantum circuit:
one can compute each term in different run of quantum computers and sum up them by classical computers.
The advantage of using the ancilla is that we can compute $M$ and $V$ for exponentially increasing number of input states in terms of the number of ancilla qubits. For example, when $L=4$ one should sum up results of four runs of quantum computers without the ancilla whereas one run is necessary with the ancilla (accompanying with the drawback of the complicated quantum circuit).
We remark that the evaluation of the transition amplitude between the time-evolved states $\ket{\psi_l(\vec{\theta})}_s= U_s (\vec{\theta})\ket{\psi_l}_s$ requires some ancilla qubits in general such as Fig.~\ref{fig:efficient circuit}.

\begin{figure}
\centering
\includegraphics[width=.25\textwidth]{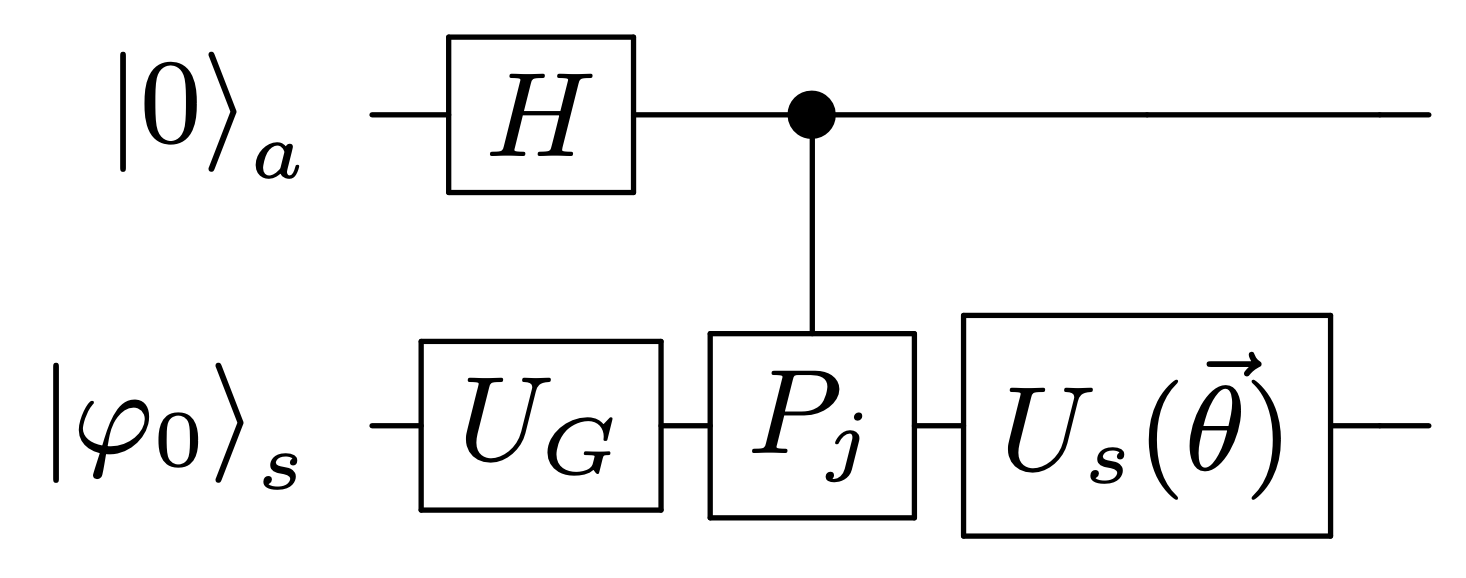}
\caption{The ansatz quantum circuit for the VQS algorithm to construct the variational unitary gate $U_s(\vec{\theta}(t))$ which approximates the time evolution operator $U(t)=e^{-iHt}$ for $\ket{\mr{G}}$ and $P_j\ket{\mr{G}}$.
Here, we assume that the ground state $\ket{\mr{G}}$ is obtained as $\ket{\mr{G}}=U_G\ket{\varphi_0}$.}
\label{fig:ansatz for GF}
\end{figure}

In the case of calculation of the Green's function, the initial states are $\ket{\psi_0} = \ket{\mr{G}}$ and $\ket{\psi_1} = P_j\ket{\mr{G}}$.
The ansatz quantum state~\eqref{eq:Ansatz for geneal VQS} for the VQS algorithm can be constructed by a quantum circuit shown in Fig.~\ref{fig:ansatz for GF}.

To sum up, the calculation of each term in Eq.~\eqref{Eq:decomposedGF}, and consequently the Green's function, with the VQS algorithm proceeds as follows:
\begin{itemize}
 \item[0.] Prepare the approximate ground state of a given Hamiltonian $H$ by conventional methods on near-term quantum computers, such as the VQE~\cite{peruzzo2014variational,kandala2017hardware,moll2018quantum,mcclean2016theory}.
 We denote the ground state as $\ket{\mr{G}}=U_G\ket{\varphi_0}$.
 \item[1.] Construct the variational quantum circuit $U(\vec{\theta})$ which approximates the time evolution $e^{-iHt}$ for two initial states $\ket{\mr{G}}$ and $P_j\ket{\mr{G}}$. The VQS algorithm with the circuit shown in Fig.~\ref{fig:ansatz for GF} will find such $U(\vec{\theta})$.  
 \item[2.] Evaluate $\bra{\mr{G}}U(\vec{\theta})^\dag P_i U(\vec{\theta}) P_j \ket{\mr{G}}$ by the quantum circuit shown in Fig.\ref{fig:efficient circuit}.
\end{itemize}

Finally, we present a detailed resource estimation about the number of required distinct runs of the quantum circuits for this algorithm in Appendix~\ref{appendix:resource}. 

\section{Computation of Green's function with excited-states search algorithm \label{sec:SSVQE}}
In this section, we describe another method to compute the Green's function of a given quantum system.
We compute the energy eigenstates and transition amplitudes of fermion operators by the algorithm based on the SSVQE method~\cite{nakanishi2018subspace} and the MCVQE method~\cite{parrish2019quantum}, and take advantage of the Lehmann representation of the spectral function~\eqref{Eq:Lehmann}.
We discuss two types of algorithms for calculating the excited states and the transition amplitudes.
The first one is based on the SSVQE algorithm with different weights where one obtains the excited states directly on quantum computers, while the second one is based on the SSVQE algorithm with identical weights where some classical post-processing after the use of quantum computers are required.
The computation of the first algorithm is simpler than that of the second one, but the convergence of the algorithm is better for the second one in general.
We note that the essential part of the algorithm described in this section is already discussed in Refs.~\cite{parrish2019quantum,nakanishi2018subspace}, so our contribution will be application of it for calculation of the Green's function.

\subsection{Computation by SSVQE with different weights}
Let us consider finding $K$ smallest eigenenergies and eigenstates of a given Hamiltonian.
The SSVQE algorithm with different weights finds the variational quantum circuit $U(\vec{\theta}^*)$ which makes input orthonormal states $\{ \ket{\psi_j}_{j=0}^{K-1} \}$ into approximate eigenstates of $H$, $\ket{\td{E}_j} := U(\vec{\theta}^*)\ket{\psi_j}$.
The approximate eigenenergies are obtained as $\td{E}_j := \bra{\td{E}_j} H \ket{\td{E}_j} = \bra{\psi_j} U(\vec{\theta}^*)^\dag H U(\vec{\theta}^*)\ket{\psi_j}$.
The algorithm performs this task by minimizing the cost function
\begin{align}
 \mathcal{C}_0(\vec{\theta})= \sum_{j=0}^{K-1} w_j \bra{\psi_j}U(\vec{\theta})^\dag H U(\vec{\theta})\ket{\psi_j}, 
\end{align}
with respect to the parameters $\vec{\theta},$ where $w_0>...>w_{K-1}>0$ are weights which assure the approximate eigenstates $\{\td{E}_j \}_j$ have the ascending order, $\td{E}_0 \leq ... \leq \td{E}_{K-1}$.
After convergence of the classical minimization for $\mathcal{C}_0(\vec{\theta})$, one obtains optimal parameters $\vec{\theta}^*$ and can compute $\{ \td{E}_j, \ket{\td{E}_j}\}_j$.

To compute the Lehmann representation of the spectral function~\eqref{Eq:Lehmann}, we also need the transition amplitude of the fermions $c_{k,\uparrow}, c_{k,\uparrow}^\dag$, such as $\braket{\td{E_j}|c_{k,\uparrow}|\td{E_k}}$.
In general, the evaluation of the transition amplitude between different quantum states needs a complicated quantum circuit,
but in this case the evaluation will be done in a simple way due to the fact that $\ket{\td{E}_j}$'s are created from the same unitary gate $U(\vec{\theta}^*)$.
Specifically, if we can easily make superpositions of the input states, $\ket{\psi_j}$ and $\ket{\psi_k}$, $\braket{\td{E_j}|c_{k,\uparrow}|\td{E_k}}$ can be evaluated by simply taking the expectation value of $c_{k,\uparrow}$ for several superpositions.
To see this, first we map $c_{k \uparrow}$ into qubit operators like Eq.~\eqref{Eq:decomposedGF} and decompose it into real part and imaginary part, $c_{k \uparrow}=A_k+iB_k$, where $A_k$ and $B_k$ are hermitian operators.
Then we have
\begin{equation}
\begin{aligned}
\bra{\psi_{j''}} U(\vec{\theta}^*)^\dag c_{k \uparrow} ~ U(\vec{\theta}^*)\ket{\psi_{j'}} = \bra{\psi_{j''}} U(\vec{\theta}^*)^\dag A_k  U(\vec{\theta}^*)\ket{\psi_{j'}} \\ 
+i \bra{\psi_{j''}} U(\vec{\theta}^*)^\dag B_k  U(\vec{\theta}^*)\ket{\psi_{j'}}.
\end{aligned}
\end{equation}
Each term can be evaluated by using $\ket{\psi^{\pm}_{j',j''}}= U(\vec{\theta}^*)(\ket{\psi_{j'}} \pm \ket{\psi_{j''}})/\sqrt{2}$ and $\ket{\psi^{i \pm}_{j',j''}} = U(\vec{\theta}^*)(\ket{\psi_{j'}} \pm i \ket{\psi_{j''}})/\sqrt{2}$ as
\begin{equation}
\begin{aligned}
\mathrm{Re}\bigg(\bra{\psi_{j''}} U(\vec{\theta}^*)^\dag A_k  U(\vec{\theta}^*)\ket{\psi_{j'}} \bigg) &= \bra{\psi^{+}_{j',j''}}A_k \ket{\psi^{+}_{j',j''}} \\ &-\bra{\psi^{-}_{j',j''}} A_k \ket{\psi^{-}_{j',j''}} \\
\mathrm{Im}\bigg(\bra{\psi_{j''}} U(\vec{\theta}^*)^\dag A_k  U(\vec{\theta}^*)\ket{\psi_{j'}} \bigg)&=
\bra{\psi^{i+}_{j',j''}}A_k \ket{\psi^{i+}_{j',j''}} \\ 
&-\bra{\psi^{i-}_{j',j''}}A_k \ket{\psi^{i-}_{j',j''}},
\end{aligned}
\end{equation}
and similar equations for the $B_k$ term.

In typical situations, the input states are taken as simple states, e.g., computational basis, so preparing superpositions of them on quantum computers is not so difficult.
Therefore, by substituting eigenenergies of Eq.~(\ref{Eq:Lehmann}) with $\td{E}_j$ and the transition amplitudes with $\braket{\td{E_0}|c_{k,\uparrow}|\td{E_k}}$, we can evaluate the spectral function, and the Green's function accordingly. 

\subsection{Computation by SSVQE with identical weights}
Next, we introduce another type of algorithms to obtain excited states and the transition amplitude between them.
This algorithm combines the SSVQE algorithm with identical weights and the quantum subspace expansion method~\cite{PhysRevA.95.042308,colless2018computation}, and is essentially the same as the MCVQE algorithm~\cite{parrish2019quantum}.

The procedure of the algorithm is as follows.
First, we prepare orthonormal  input states $\{ \ket{\psi_j}_{j=0}^{K-1} \}$, which are simple and easy to realize a superposition of them on quantum computers.
Then we minimise the cost function
\begin{align}
\mathcal{C}_1(\vec{\theta})= \sum_{j=0}^{K-1} \bra{\psi_j}U^\dag(\vec{\theta})H U(\vec{\theta})\ket{\psi_j}, 
\end{align}
with respect to parameters $\vec{\theta}$, where $U(\vec{\theta})$ is the ansatz quantum circuit.
After the optimisation, the subspace spanned by $\{U(\vec{\theta}^*)\ket{\psi_j}\}_{j=0}^{K-1}$ will be close to that spanned by the true $K$ eigenstates $ \{\ket{E_j} \}_{j=0}^{K-1}$, where $\vec{\theta}^*$ is the parameters after optimisation.
At this stage, $U(\vec{\theta}^*)\ket{\psi_j}$ is generally the superposition of the excited states $\{\ket{E_j} \}_{j=0}^{K-1}$ and $\bra{\psi_j} U(\vec{\theta}^*)^\dag H U(\vec{\theta}^*)\ket{\psi_j}$ is not a good approximation to the true eigenvalue $E_j$.
Therefore, to obtain the eigenstates and eigenvalues of $H$, we solve the eigenvalue problem within the subspace spanned by $\{U(\vec{\theta}^*)\ket{\psi_j}\}_{j=0}^{K-1}$,
\begin{align}
 \mathcal{H} \mathcal{V}= \mathcal{V} \mathcal{E}, 
\label{Eq:HVeq}
\end{align}
where $\mathcal{H}_{i,j}=\bra{\psi_i}U^\dag(\vec{\theta}^*)H U(\vec{\theta}^*)\ket{\psi_j}$,
$\mathcal{V}$ is $K\times K$ matrix containing eigenvectors as its columns, and $\mathcal{E}$ is a diagonal matrix whose diagonal elements are eigenvalues.
The approximate $m$-th excited state $\ket{\td{E}'_m}$ is expressed as 
\begin{align}
 \ket{\td{E}'_m}=\sum_j \mathcal{V}_{j, m} U(\vec{\theta}^*)\ket{\psi_j},
\end{align}
and the approximate eigenenergies appear as $\td{E}'_j = \mathcal{E}_{j,j}$.

The transition amplitude $C_{m,n}^{(k)}=\bra{\td{E}'_m} c_{k \uparrow} \ket{\td{E}'_n}$ can be computed as 
\begin{align}
C_{m,n}^{(k)}= \sum_{j', j''} \mathcal{V}_{j'', m}^* \mathcal{V}_{j', n} \bra{\psi_{j''}} U(\vec{\theta}^*)^\dag c_{k \uparrow} ~ U(\vec{\theta}^*)\ket{\psi_{j'}}.
\end{align}
The quantity $\bra{\psi_{j''}} U(\vec{\theta}^*)^\dag c_{k \uparrow} ~ U(\vec{\theta}^*)\ket{\psi_{j'}}$ can be evaluated in the way described in the previous subsection.
Thus, we can compute the transition matrix $C^{(k)}_{m,n}$, and evaluate the spectral function by Eq.~(\ref{Eq:Lehmann}) and the Green's function.

\section{Numerical demonstration \label{sec:Numerics}}
In this section, we numerically demonstrate our proposed methods to calculate the Green's function and the spectral function at zero temperature.
We consider the Fermi-Hubbard model of two sites with the particle-hole symmetry,
\begin{align}
  H &=& -t \sum_{\sigma=\uparrow,\downarrow} \left( c_{1,\sigma}^\dagger c_{2,\sigma} + \rm{h.c.} \right)
 + U \sum_{i=1}^2 c_{i,\uparrow}^\dagger c_{i,\uparrow} c_{i,\downarrow}^\dagger c_{i,\downarrow} \nonumber \\
 && -\frac{U}{2} \sum_{i=1,2,\sigma=\uparrow,\downarrow} c_{i,\sigma}^\dag c_{i,\sigma}
 \label{Hubbard_def}
\end{align}
where $t$ is a parameter characterizing hopping between the sites and $U$ denotes the strength of the on-site Coulomb repulsion~\cite{Gutzwiller1963,Kanamori1963,hubbard1963electron}.
We set the hopping parameter $t=1$ throughout this paper.
We simulate two proposed protocols in Secs.~\ref{sec:VQS} and~\ref{sec:SSVQE} by classical computers with the fast quantum circuit simulation library Qulacs~\cite{qulacs_2018}.
We use the Jordan-Wigner transformation~\cite{Jordan1928} to map the fermionic Hamiltonian~\eqref{Hubbard_def} into the qubit one with four qubits by using the library OpenFermion~\cite{mcclean2017openfermion}.

\subsection{Numerical simulation of the method based on variational quantum simulation}
\begin{figure}
    \centering
    \includegraphics[width=.35\textwidth]{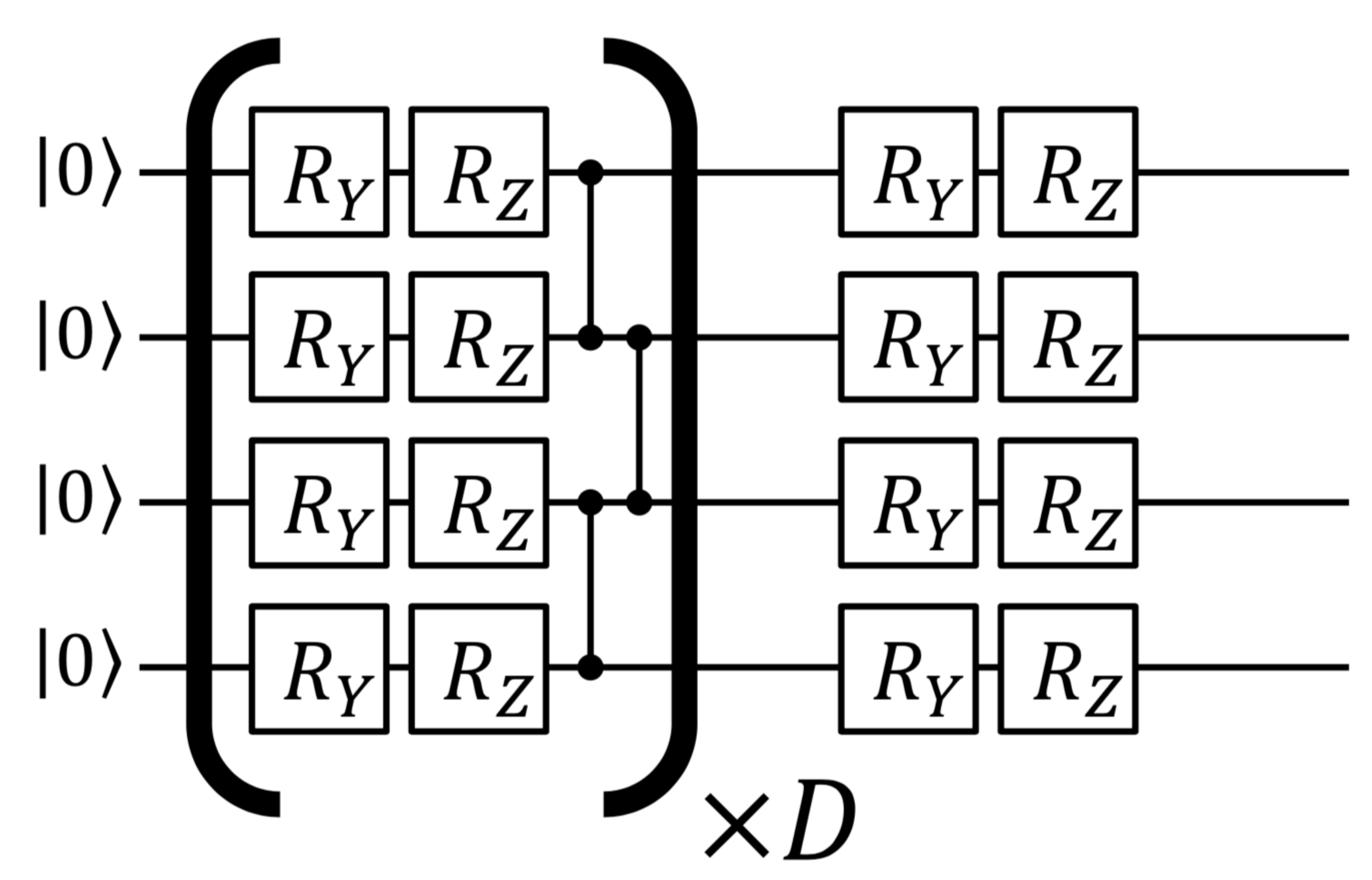}
    \caption{Hardware-efficient-type ansatz used to obtain the ground state of the model~\eqref{Hubbard_def} for the VQS algorithm.
    Each rotational gate $R_Y(\theta)=e^{i\theta Y/2}, R_Z(\theta')=e^{i\theta' Z/2}$
    \label{fig:HEA} has a parameter angle and $D$ denotes the depth of the ansatz.}
\end{figure}

\begin{figure*}
     \includegraphics[width=.3\textwidth, trim=20 0 20 20]{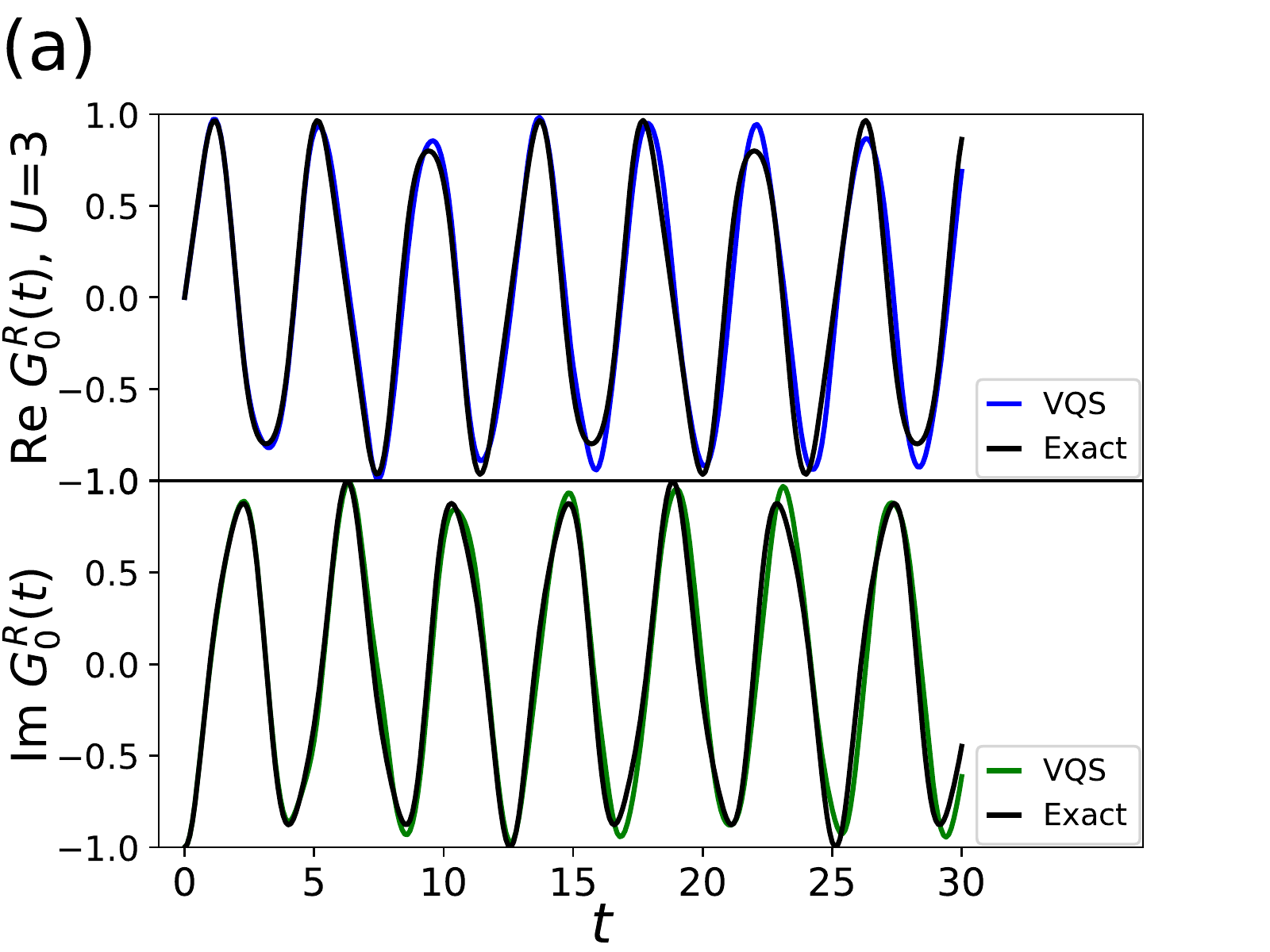}
     \includegraphics[width=.3\textwidth, trim=20 0 20 20]{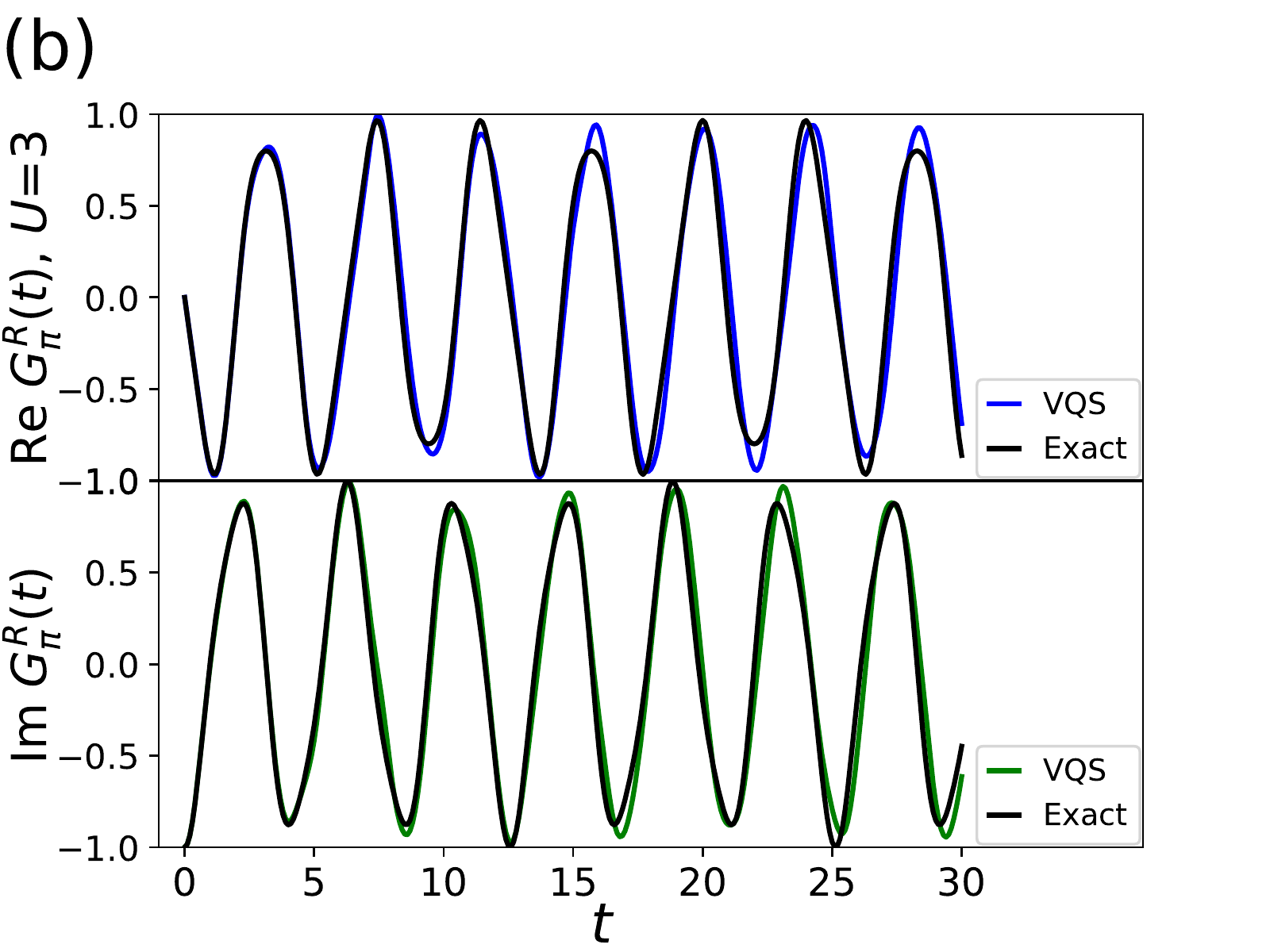}
     \includegraphics[width=.3\textwidth, trim=20 0 20 20]{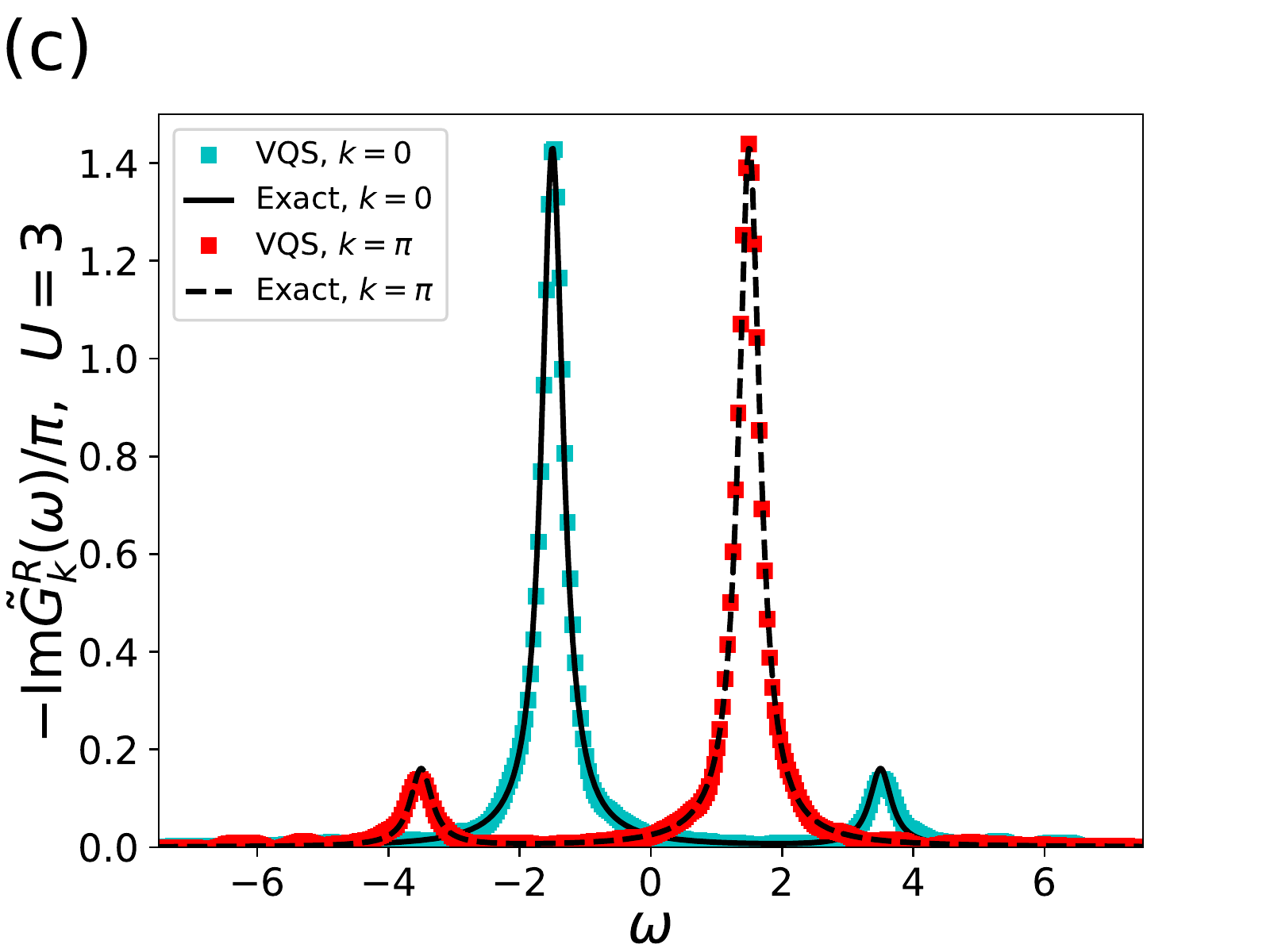} \\
     \includegraphics[width=.3\textwidth, trim=20 20 20 20]{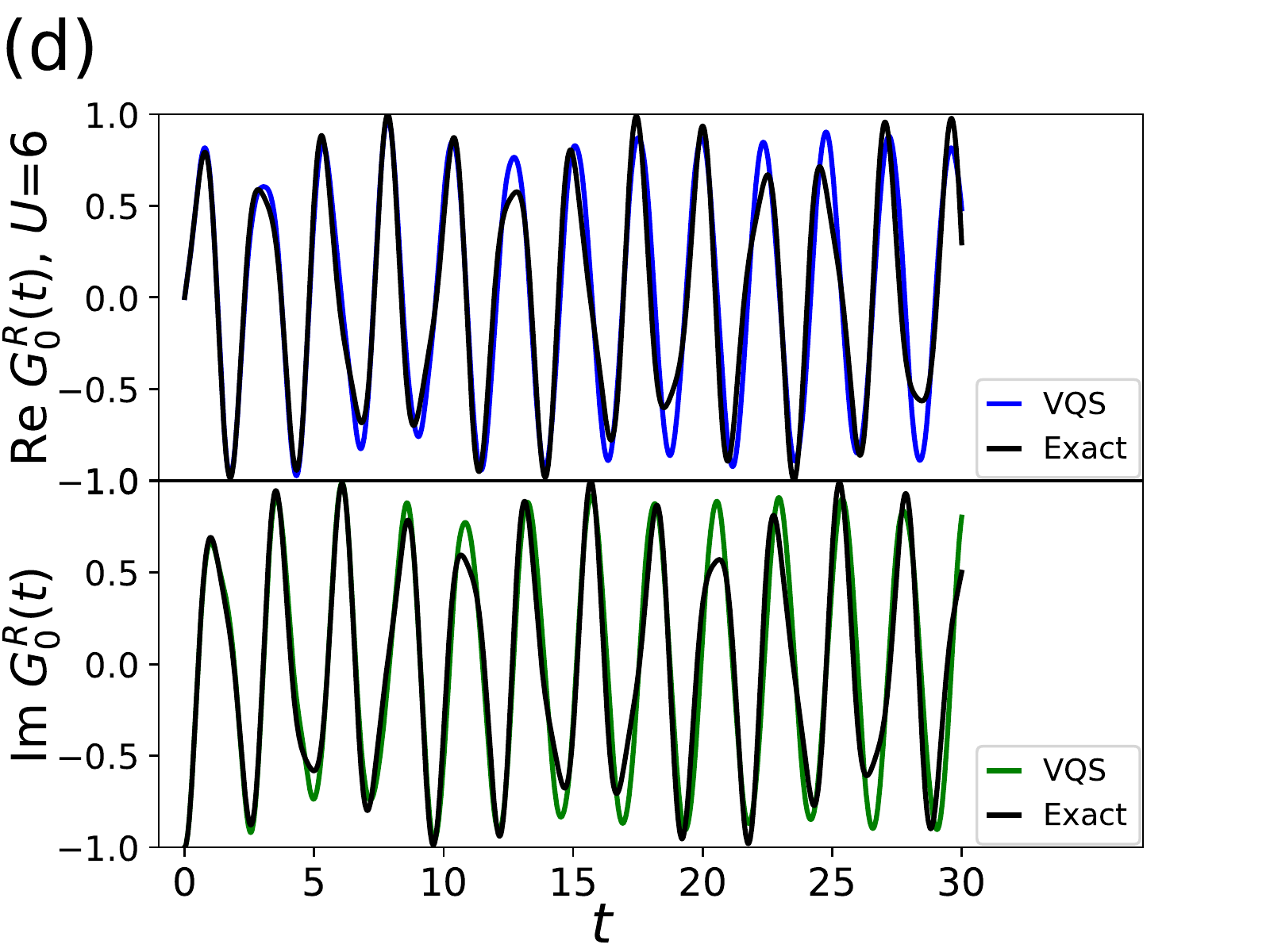}
     \includegraphics[width=.3\textwidth, trim=20 20 20 20]{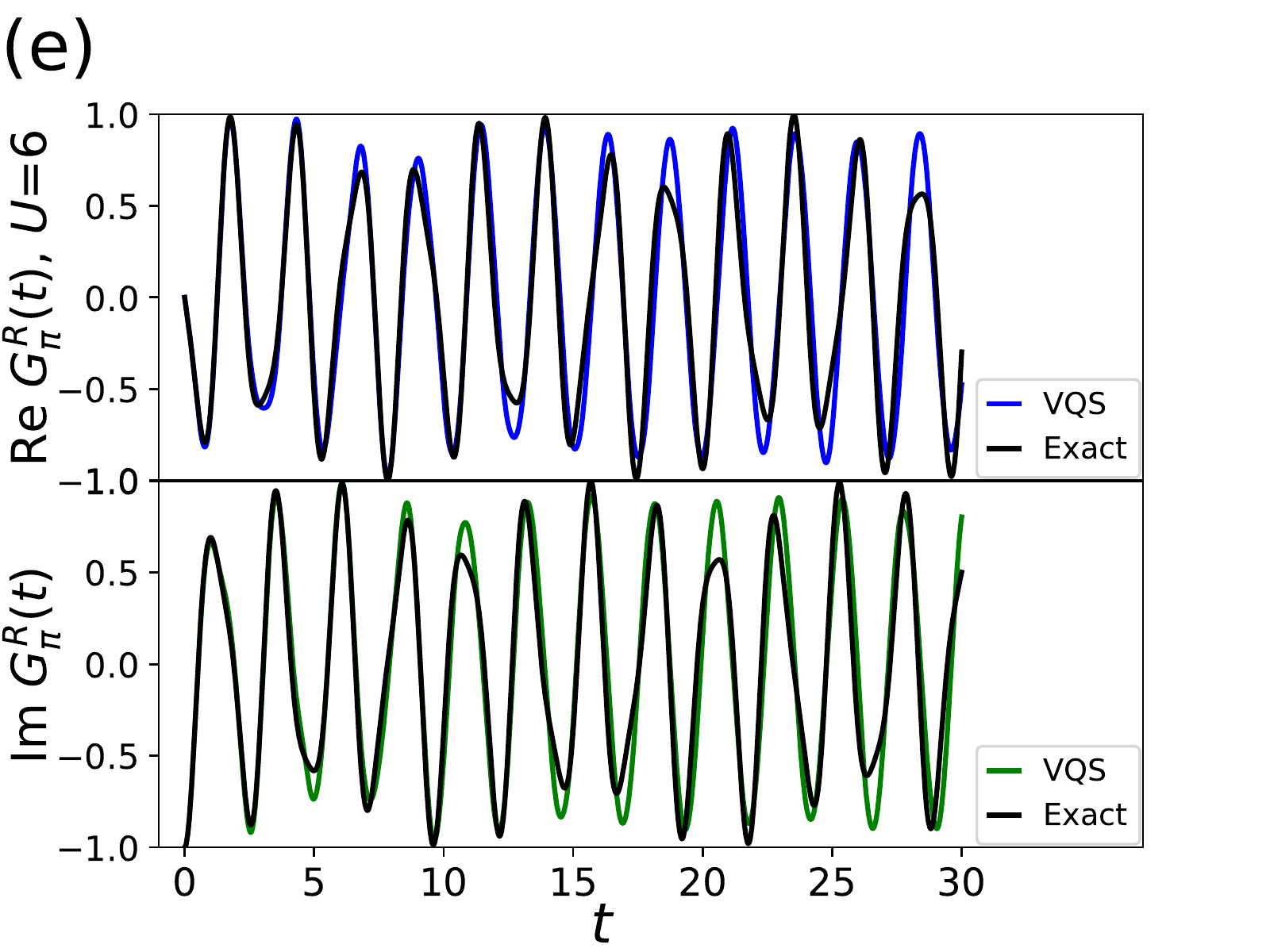}
     \includegraphics[width=.3\textwidth, trim=20 20 20 20]{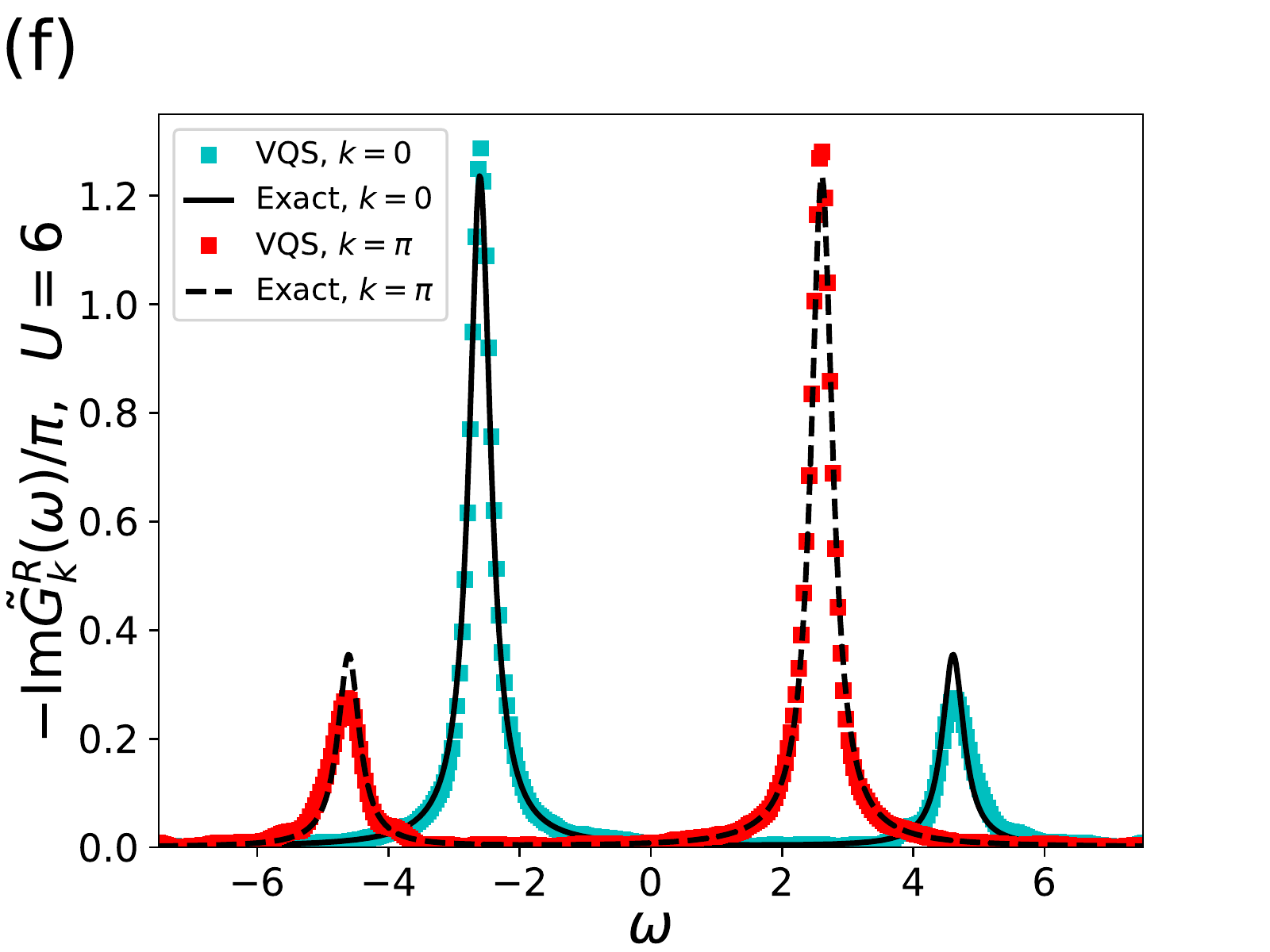}
   \caption{\label{fig:VQS result} Numerical simulation of the VQS algorithm to compute the Green's function in real time $G^R_k(t)$ (a, b, d, e) and the spectral function (c, f) for the model~\eqref{Hubbard_def} of $U=3$ (a-c) and $U=6$ (d-f). 
   The time step is taken as $dt=0.1 (0.03)$ for $U=3 (6)$.
   The exact spectral function is calculated by the exact dynamics of the Green's function in real time from $t=0$ to $t=100$ with step $dt=0.1$.
   We take $\eta = 0.2$ for the calculation of the spectral functions.}
\end{figure*}

We calculate the real-time Green's function of the model~\eqref{Hubbard_def} at zero temperature by using the method described in Sec.~\ref{sec:VQS}.
First, we prepare the ground state of the model~\eqref{Hubbard_def} by the standard VQE algorithm with a hardware-efficient-type ansatz~\cite{kandala2017hardware} depicted in Fig.~\ref{fig:HEA}.
Then, we perform the VQS algorithm.
As an ansatz quantum state, we adopt the so-called variational Hamiltonian ansatz~\cite{Wecker2015, Reiner_2019} inspired by the Suzuki-Trotter decomposition of the time evolution operator $e^{-iHt}$.
The variational Hamiltonian ansatz is defined through the qubit representation of the Hamiltonian,
\begin{align}
  H_\mr{qubit} = \sum_{m} c_m P_m,
 \label{Hubbard_def_qubit}
\end{align}
where $P_m$ is a (multi-qubit) Pauli matrix and $c_m$ is a coefficient.
An ansatz state for the variational Hamiltonian ansatz is given by $\ket{\psi(\theta)} = U_{VHA}(\theta) \ket{\varphi_0}$, where
\begin{align}
  U_{VHA}(\{ \theta_m^{(d)} \}_{m,d}) = \prod_{d=1}^{n_d} \left( \prod_m \exp\left( i \theta_m^{(d)} P_m \right) \right),
 \label{HamAnsatz_def}
\end{align}
and $n_d$ denotes the depth of the ansatz.
We note that we remove the identity operator from Eq.~\eqref{Hubbard_def_qubit} when constructing the ansatz.
If the qubit representation of the Hamiltonian~\eqref{Hubbard_def_qubit} has $N_P$ terms (except for the identity operator), the number of parameters of the ansatz will be $N_P n_d$.
In our simulation, $N_P=6$ and we choose $n_d=8$, so there are 48 parameters in the parametrized quantum circuit whereas general unitary operators on the system have $(2^4)^2 = 256$ parameters. 
Further details on numerical calculations are described in Appendix~\ref{App: details}.

The result is shown in Fig.~\ref{fig:VQS result}.
The VQS algorithm nicely reproduces the exact dynamics (the panels in left and center columns), and the spectral function (right columns).
These figures illustrate the possibility of the VQS algorithm proposed in this study to calculate the Green's function.
In section~\ref{app:dependence_n_d}, dependence of the results on the depth of the ansatz is analyzed.
Furthermore, numerical simulations for $n_d=4$ and the four-site Fermi-Hubbard model are presented in Appendix~\ref{additional}. One can see that numerical results for $n_d=4$ have almost the identical performance compared to the case of $n_d=8$; therefore depending on strength of physical noises, the choice of $n_d=4$ may be recommended to avoid accumulation of physical errors.

\subsection{Numerical simulation of the method based on excited-state search}
\begin{figure}
\includegraphics[width=.5\textwidth]{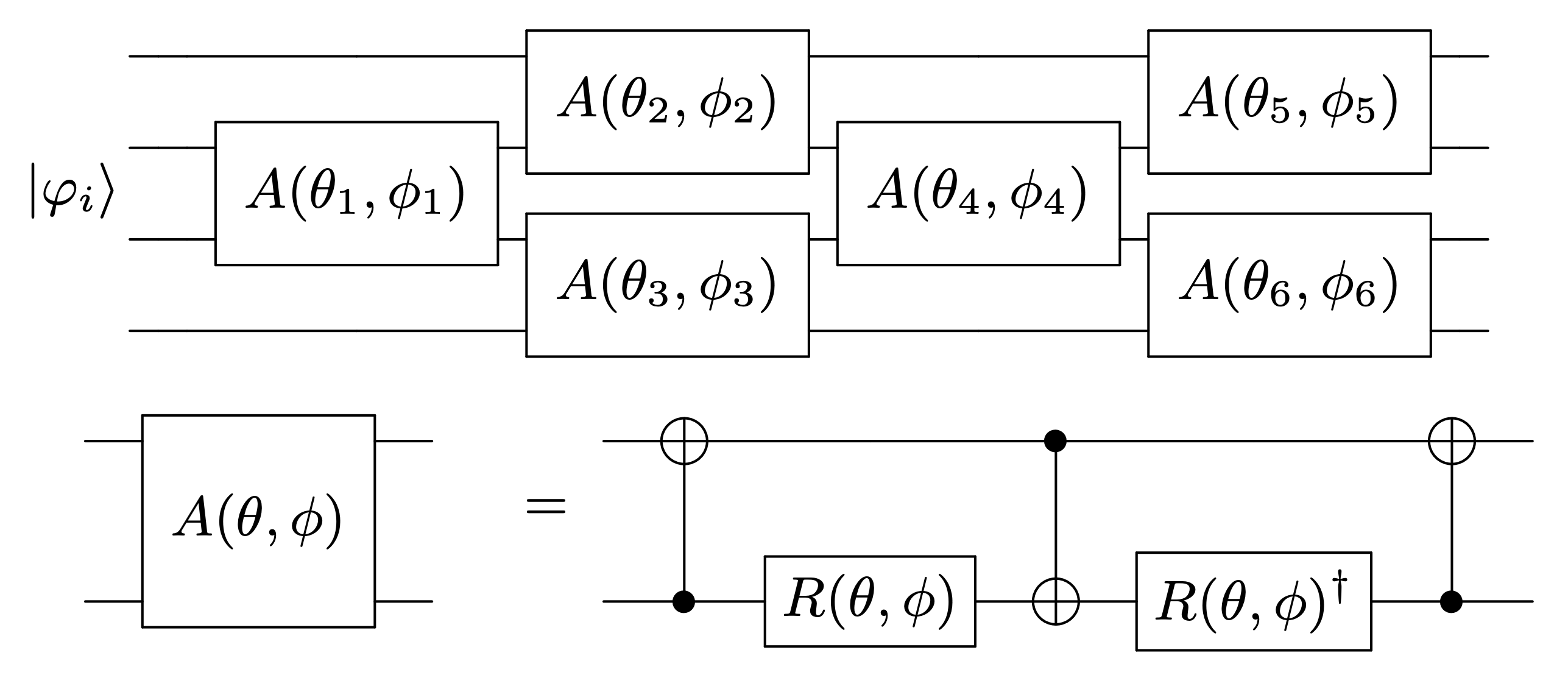}
\caption{(top) Definition of the symmetry-preserving ansatz~\cite{Gard2019} used for the demonstration of the SSVQE algorithm with identical weight.
(bottom) Definition of the $A$ gate. Here, $R(\theta,\phi)$ is defined as $R(\theta,\phi)=R_Y(\theta+\pi/2)R_Z(\phi+\pi)$, where $R_Y(\theta)=e^{i\theta Y/2}$ and $R_Z(\phi)=e^{i\phi Z/2}$.
\label{fig:SP ansatz} }
\end{figure}

\begin{figure*}
     \includegraphics[width=.45\textwidth]{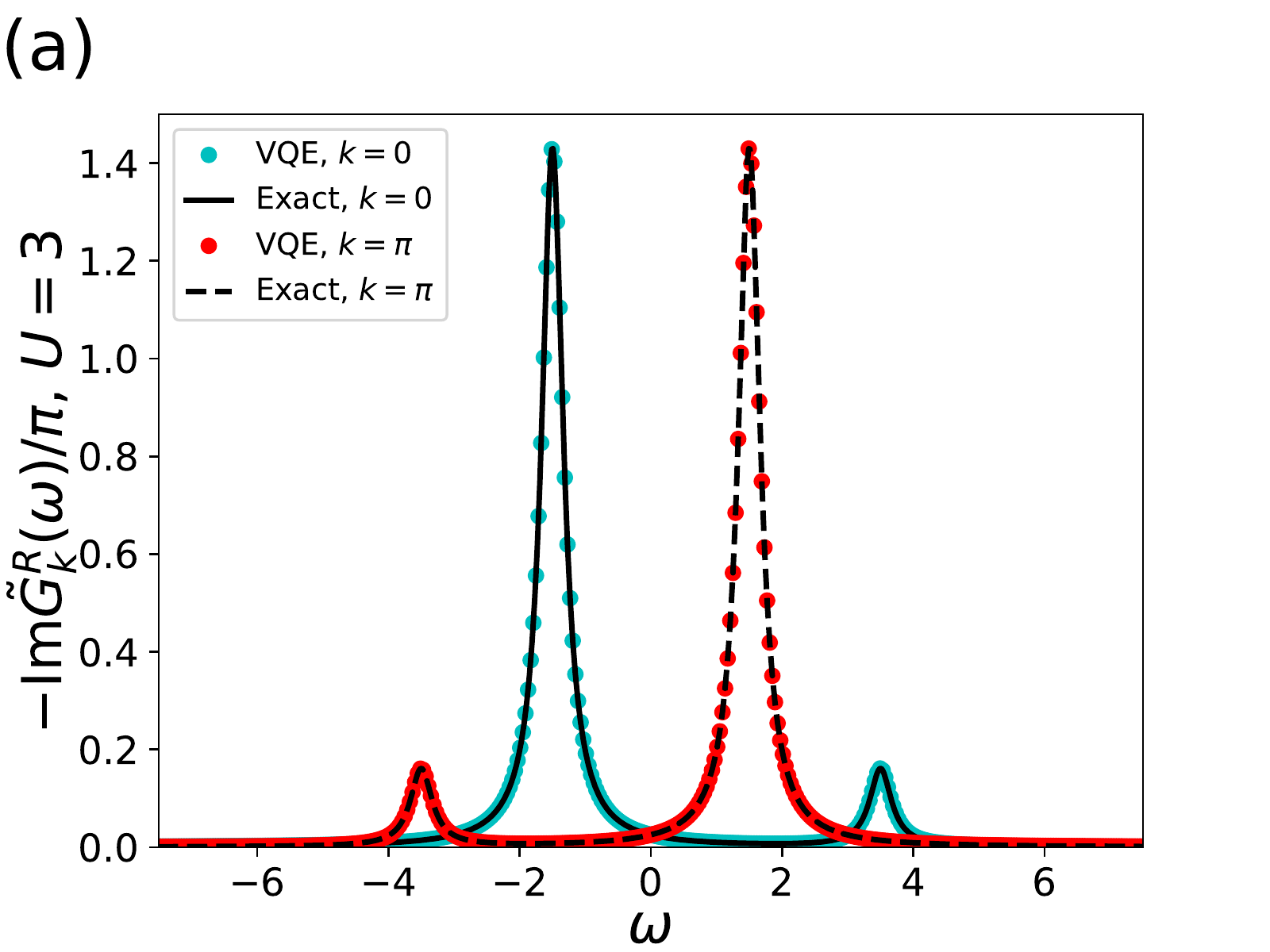}
     \includegraphics[width=.45\textwidth]{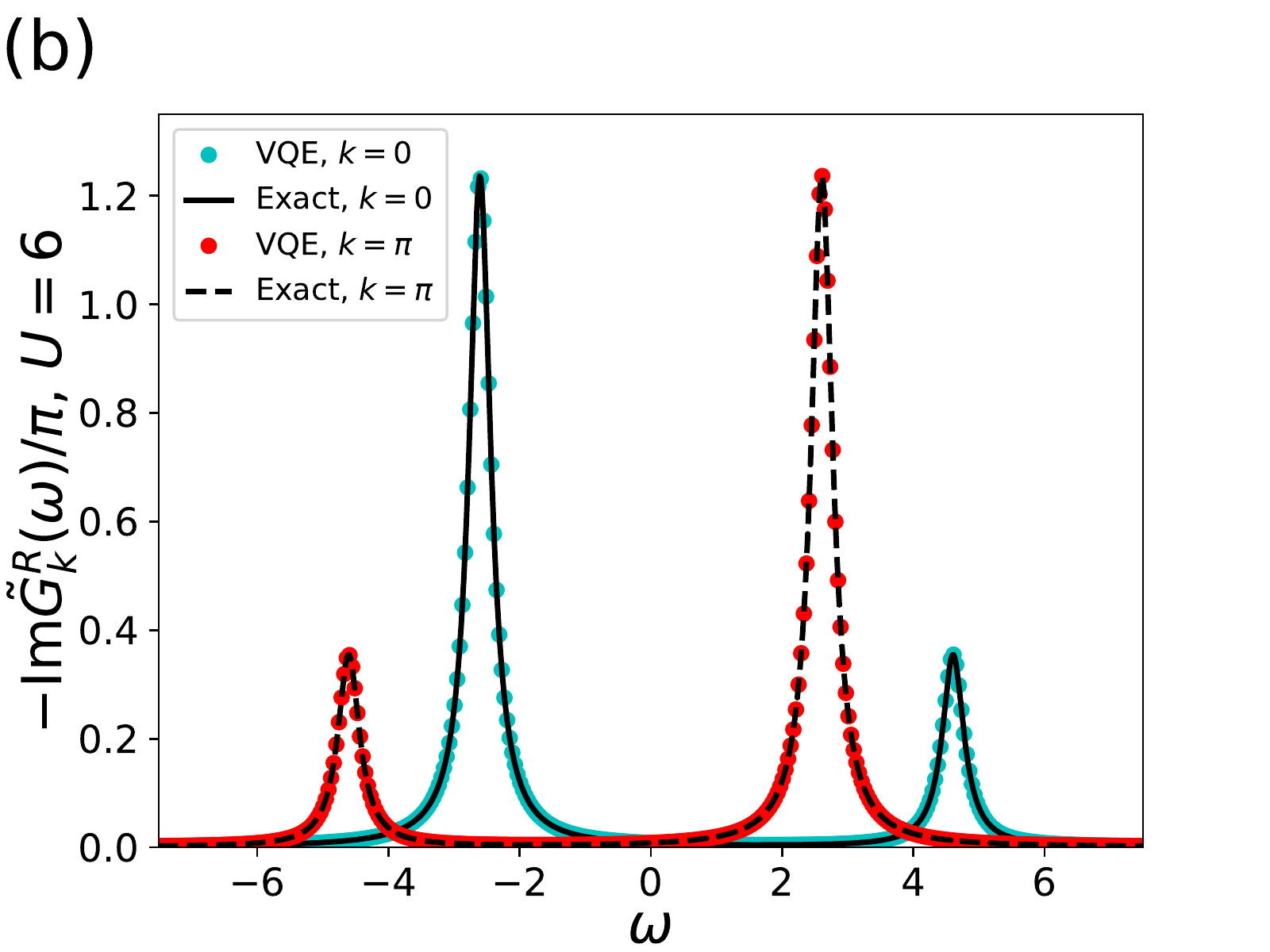}
   \caption{\label{fig:SSVQE result} Result of numerical simulation of the method described in Sec.~\ref{sec:SSVQE} for calculating the spectral function of the model~\eqref{Hubbard_def} at $U=3$ (a) and $U=6$ (b). }
\end{figure*}

Next, we numerically simulate the method described in Sec.~\ref{sec:SSVQE}.
We adopt the symmetry-preserving ansatz~\cite{Gard2019} drawn in Fig.~\ref{fig:SP ansatz}, which preserves the total number of particles in the system.
We use the SSVQE algorithm with identical weight and calculate five energy eigenstates of the model~\eqref{Hubbard_def}.
As an input states, we simply choose the computational basis states with desired particle number: to calculate the particle (hole) part of the spectrum function, we choose $\ket{\varphi_0}=\ket{0011}$ for the ground state,  $\ket{\varphi_1}=\ket{0001}, \ket{\varphi_2}=\ket{0010}, \ket{\varphi_3}=\ket{0100}, \ket{\varphi_4}=\ket{1000}$ ($\ket{\varphi_1}=\ket{0111}, \ket{\varphi_2}=\ket{1011}, \ket{\varphi_3}=\ket{1101}, \ket{\varphi_4}=\ket{1110}$) for the excited states. 

Figure~\ref{fig:SSVQE result} shows the result of numerical simulation.
The SSVQE algorithm almost perfectly reproduces the exact result obtained by exact diagonalization.

\section{ Dependence of accuracy of the variational quantum simulation on depth of the ansatz }  
\label{app:dependence_n_d}
In this section, we provide a systematic analysis on how the accuracy of the numerical simulations of the variational quantum simulation (VQS) in Sec.~\ref{sec:Numerics} depends on the depth $n_d$ of the Hamiltonian ansatz (Eq.~\eqref{HamAnsatz_def}).
We run numerical simulations for $n_d=4,5,\cdots,10$ under the same conditions as in Fig.~\ref{fig:VQS result}.

\begin{figure}[h!]
     \includegraphics[width=.5\textwidth]{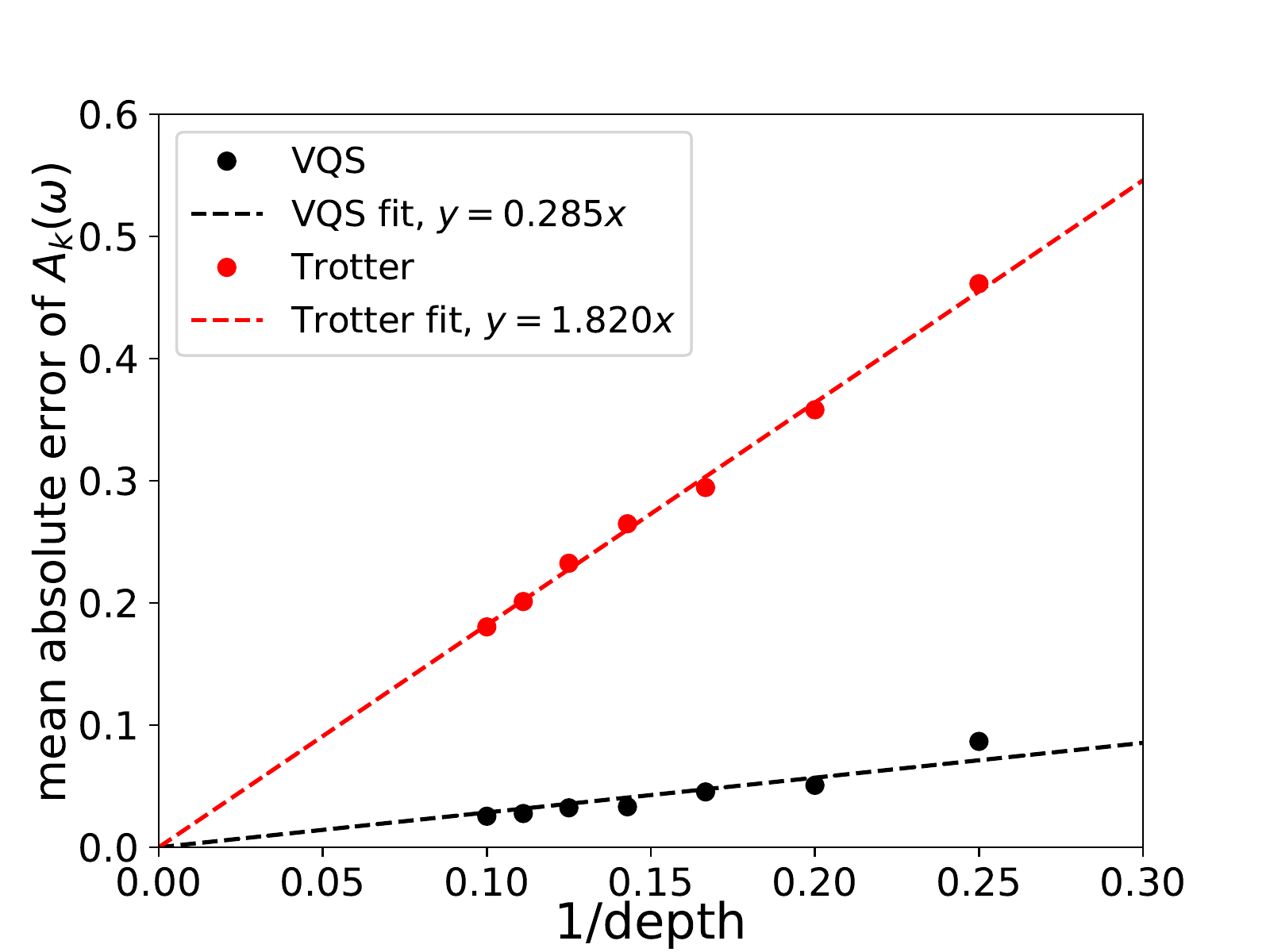}
   \caption{\label{appfig:inverse_residue}
   The MAEs (Eq.~\eqref{appeq:MAE}) calculated for the spectrum function obtained by the VQS (black dots) and the Suzuki-Trotter decomposition of the time evolution operator, Eq.~\eqref{appeq:STdecomp} (red dots), for $k=\pi$ and $U=3$.
   Dotted lines represent fittings of the data by a linear function $y=ax$, where the slope $a$ is a parameter to be optimized. The values of the slope for both cases are shown in the legend.}
\end{figure}

To see the accuracy of the simulations quantitatively, we calculate the mean absolute error (MAE) of the spectrum function at $k=\pi$ in the region of $\omega \in [-5, 5]$,
\begin{equation} \label{appeq:MAE}
\begin{aligned}
  \Delta_E (k) = \frac{1}{2N_\omega + 1}&\sum_{n=-N_\omega}^{N_\omega}
  \bigg| A^\mr{exact}\left(k, \omega = \frac{5n}{N_\omega}\right) \\
  &- A^\mr{VQS}\left(k, \omega = \frac{5n}{N_\omega}\right) \bigg|,
  \end{aligned}
\end{equation}
where $2N_\omega + 1$ is the total number of data points and $A^\mr{exact(VQS)}(k,\omega)$ is the spectrum function calculated by exact diagonalization (VQS).
We take $N_\omega=5000$.
Interestingly, as seen from Fig.~\ref{appfig:inverse_residue}, the MAE decreases with the inverse of the depth of the ansatz.
This dependence reminds us of the Suzuki-Trotter decomposition of the time evolution operator, i.e.,
\begin{equation} \label{appeq:STdecomp}
  U(t) = e^{-iH_\mr{qubit}t} \approx U^{\mr{Trot}}_{n_d}(t) = \left(\prod_m e^{-i c_m P_m \cdot \frac{t}{n_d}} \right)^{n_d}
\end{equation}
with the error of $O(t^2/n_d)$, where we have used the notation in Eq.~\eqref{Hubbard_def_qubit}.
In Fig.~\ref{appfig:inverse_residue}, we also show the MAE of the spectrum function calculated by the dynamics obtained from the approximate time evolution operator Eq.~\eqref{appeq:STdecomp} with the same time step and duration used for the VQS.
The MAE for this case exhibits $1/n_d$ dependence as expected, but the values of the MAE are much larger than those for the VQS; the slope of the fit of the MAE with $1/n_d$ is about six times smaller for the VQS  ($1.820/0.285\approx6.4$).
The result shown in Fig.~\ref{appfig:inverse_residue} not only provides an estimation of errors in the VQS calculation when the ansatzes with various depths are used, but also illustrates the practical advantage of employing the VQS compared with the Suzuki-Trotter decomposition of the time evolution operator which has the same-depth quantum circuit.

\section{Feasibility of quantum algorithms on near-term quantum computers \label{sec:gatecount}}
In this section, we discuss the feasibility of implementing our proposed algorithms on near-term quantum computers.
Let us consider 25 sites two-dimensional Fermi-Hubbard model on a square lattice, whose simulation requires 50 qubits and is almost intractable for classical computers.
The model is defined as
\begin{equation}
 H_{2d} = -t \sum_{\braket{i,j}, \sigma} \left( c_{i,\sigma}^\dagger c_{j,\sigma} + \rm{h.c.} \right)
 + U \sum_{i} c_{i,\uparrow}^\dagger c_{i,\uparrow} c_{i,\downarrow}^\dagger c_{i,\downarrow},
 \label{eq:2dHubbardDef}
\end{equation}
where $\braket{i,j}$ runs nearest-neighbor sites on  a square lattice and $\sigma=\uparrow,\downarrow$ denotes the spin.
Based on the argument in Ref.~\cite{cai2019resource}, using the Hamiltonian ansatz (Eq.~\eqref{HamAnsatz_def}) with the Jordan-Wigner transformation, we need $N_{\mathrm{two}}\approx 1000$ two-qubit gates per depth of the ansatz when we employ the rotation Z gate and partial swap gates as elementary gates.
Although near-term quantum computers contain inevitable noise in gate operations, the technique of quantum error mitigation ~\cite{temme2017error,endo2018practical} can suppress errors and recover noiseless results with a reasonable overhead when the error rate per gate $\epsilon_{\mathrm{gate}}$ satisfies $N_{\mathrm{gate}}\epsilon_{\mathrm{gate}} \lesssim 2$, where $N_{\mathrm{gate}}$ is the number of gates. Therefore, when we adopt the Hamiltonian ansatz of two depths as an ansatz quantum circuit for the VQS or the SSVQE, we need at least $2/2000 = 0.1 \%$ for the error rate of two-qubit gates, which has been achieved in current experimental setups~\cite{harty2014high,ballance2016high}. 
We note that the error of single-qubit gate is ignored here because it is negligible compared with that of two-qubit gate.
Only a few additional controlled operations in the VQS-based algorithm introduced in Sec.~\ref{sec:VQS} is also be neglected.
In addition, for the VQS-based algorithm, gates for the unitary $U_G$ which prepares the ground state of the system must be taken into account. The more detailed argument has been made in Appendix~\ref{appendix:gatecount}.

\section{Extension to finite temperature Green's function \label{sec:FiniteT}}
Here, we discuss the extension of our proposed methods to the Green's function at finite temperature.
For finite temperature $T>0$ or inverse temperature $\beta = 1/T < \infty$,
the retarded Green's function is defined as
\begin{equation}
\begin{aligned}
G^R_{k}(t;\beta) &= \frac{1}{Z(\beta)} \sum_n e^{-\beta E_n} G^R_{k,n}(t), \\
Z(\beta) &= \mr{Tr}\left( e^{-\beta H} \right), \\
G^R_{k,n}(t) & = -i \Theta(t) \left( \bra{E_n} e^{i H t} c_{k \uparrow} e^{-i H t} c_{k \uparrow}^\dag  \ket{E_n} \right. \\
& \left. + \bra{E_n}c_{k \uparrow}^\dag e^{i H t} c_{k \uparrow} e^{-i H t} \ket{E_n} \right), 
\label{Eq: finiteGF}
\end{aligned}
\end{equation}
where $\ket{E_n}$ and $E_n$ denote the eigenstates and eigenvalues of Hamiltonian $H$, respectively.
The corresponding spectral function is
\begin{equation}
\begin{aligned}
\td{G}^R_k(\omega;\beta) & = \int_{-\infty}^{\infty} dt \, e^{i(\omega+i\eta)t} G^R_k(t; \beta),\\
A_k(\omega; \beta) &= \frac{1}{Z(\beta)} \sum_{n,m} e^{-\beta E_m} \\
&\times \left( \frac{ |\braket{E_n|c_k^\dag|{E_m}}|^2 }{\omega+E_m-E_n+i\eta}  \right.
  \left. + \frac{ |\braket{E_n|c_k|{E_m}}|^2 }{\omega-E_m+E_n+i\eta} \right).
 \label{Eq:Lehmann_finite}
\end{aligned}
\end{equation}

\subsection{Variational quantum simulation for Green's function at finite temperature}
\begin{figure}
\includegraphics[width=.25\textwidth]{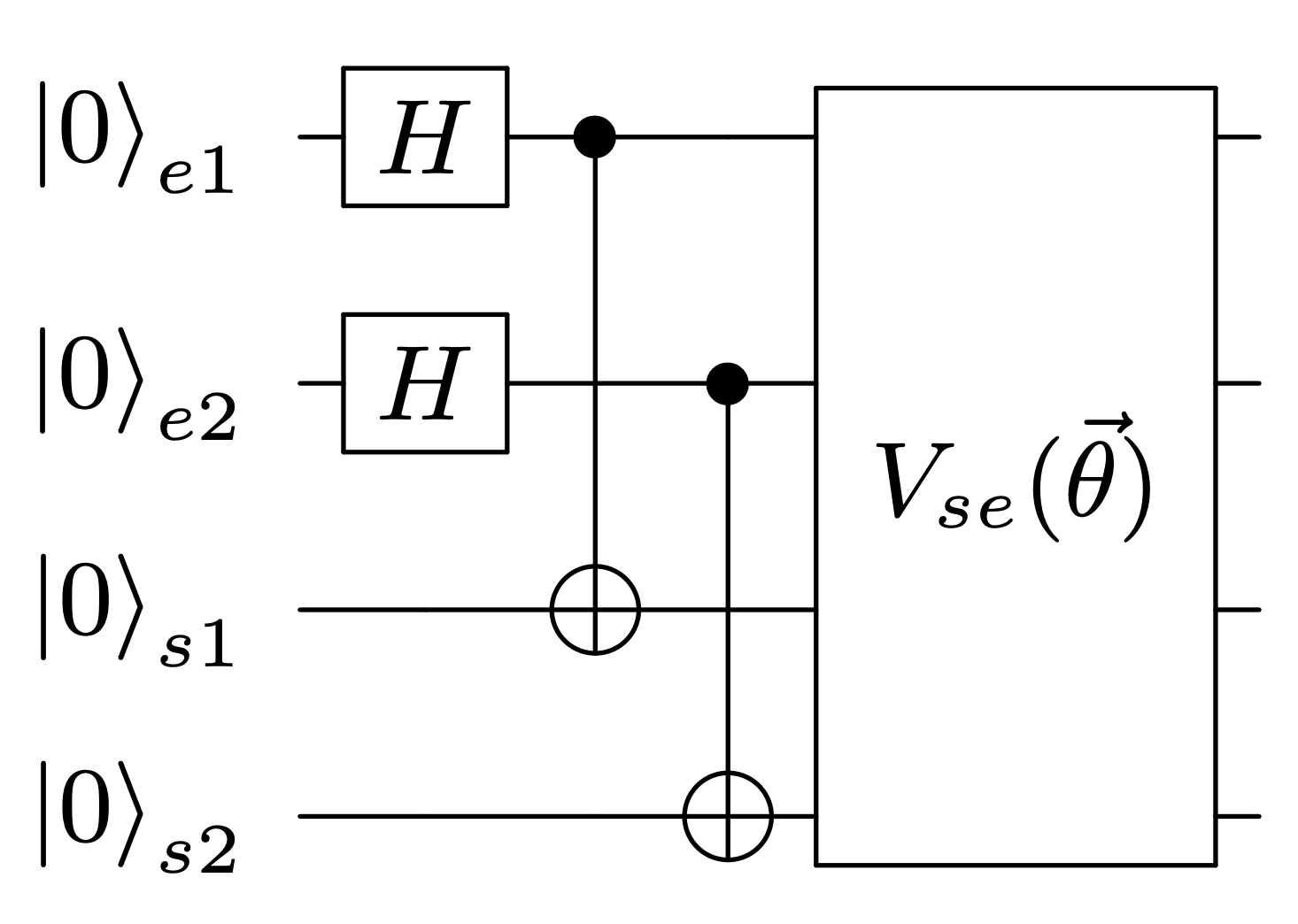}
\caption{The ansatz quantum circuit for obtaining the purified Gibbs state $\ket{\Phi(\beta)}$ for $N=2$.} 
\label{fig: GibsCircuit}
\end{figure}

\begin{figure}
\centering
\includegraphics[width=.4\textwidth]{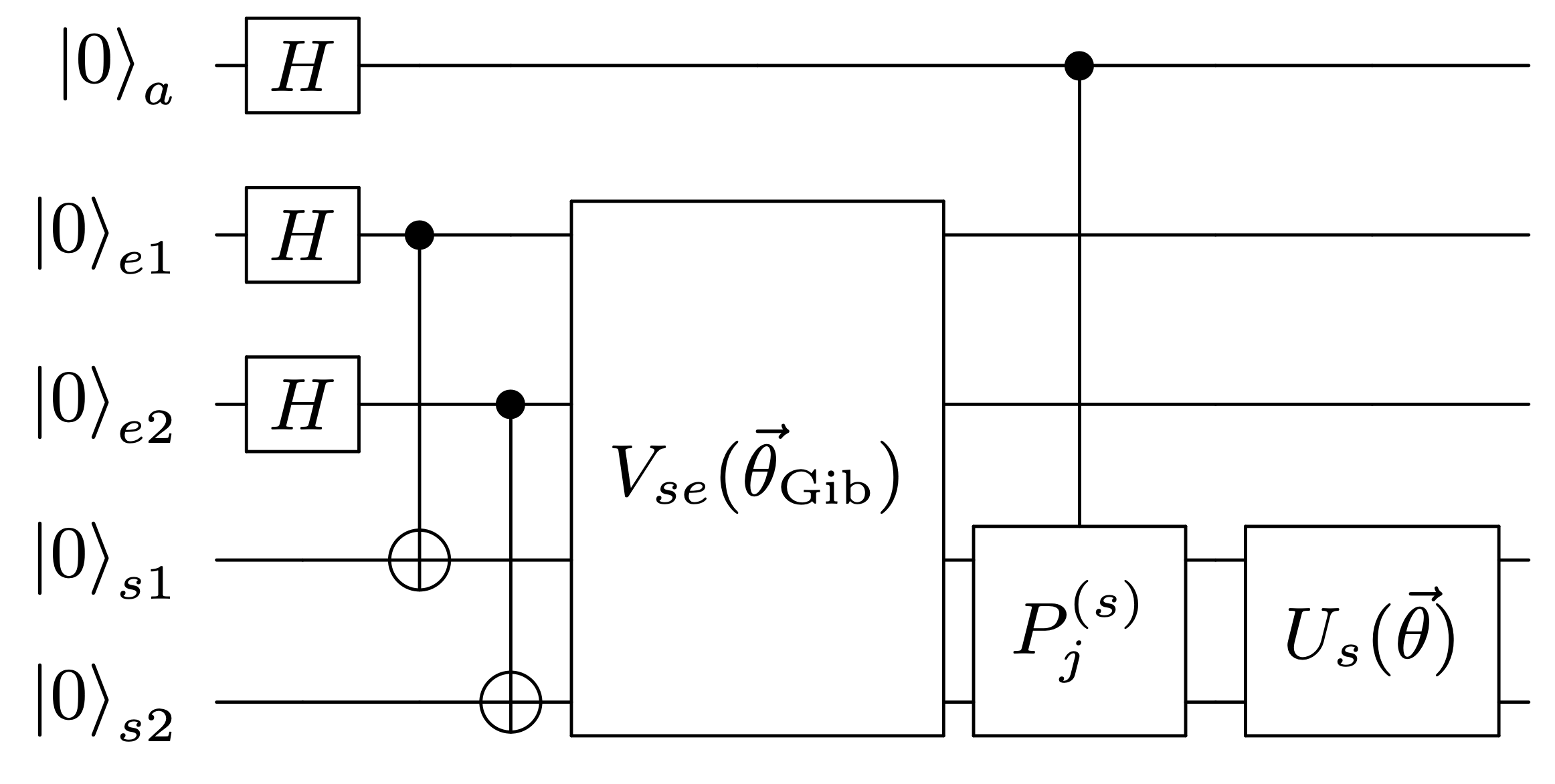}
\caption{The ansatz quantum circuit for the VQS algorithm to construct the variational unitary gate $U_s(\vec{\theta}(t))$ which approximates the time evolution operator $U(t)=e^{-iHt}$ for the purified state Gibbs state $\ket{\Phi(\beta)}=V_{se}(\vec{\theta}_{\mathrm{Gib}})\ket{\Phi_0}$ and $P_j^{(s)}\ket{\Phi(\beta)}$ simultaneously ($N=2$).} 
\label{gibbscircdynamics}
\end{figure}

Equation~\eqref{Eq: finiteGF} can be evaluated on quantum computers by combining the VQS method and the thermofield double technique which purifies the Gibbs state of the system.
The procedure is essentially the same as the finite-temperature version of the density matrix renormalization group method~\cite{Zwolak2004,Verstraete2004}.
We note that several methods based on the typically of the Hilbert space to evaluate physical quantities at finite temperature~\cite{Sugiura2012,Stoudenmire_2010,cohn2018minimal} might also be combined with the VQS algorithm.

Let us consider a $N$-qubit ``environment" system (denoted by subscript $e$) in addition to the original $N$-qubit system of interest (denoted by subscript $s$).
First We prepare a state $\ket{\Phi_0}$ which satisfies $\mr{Tr}_e \left(\ket{\Phi_0}\bra{\Phi_0} \right) = I_s$.
For example, we choose $\ket{\Phi_0} = \frac{1}{\sqrt{2^N}}\sum_{i=0}^{2^N-1} \ket{i}_s \otimes \ket{i}_e $ where $\ket{i}_{s,e}$ is the computational basis of the system and environment.
By using the method of variational imaginary time evolution introduced in Ref.~\cite{yuan2018theory}, one can obtain the state $\ket{\Phi(\beta)} \approx Z(\beta)^{-1/2} e^{-\beta H/2} \ket{\Phi_0} $.
Namely, the variational imaginary time evolution for the total system with the ``Hamiltonian" $H \otimes I_e$ and the variational quantum circuit drawn in Fig.~\ref{fig: GibsCircuit} will produce $\ket{\Phi(\beta)}$.
We note that $\ket{\Phi(\beta)}$ satisfies
\begin{align}
\mr{Tr}_e \left( \ket{\Phi(\beta)}\bra{\Phi(\beta)} \right) \approx\frac{1}{Z(\beta)} e^{-\beta H}.
\end{align}

Next, we perform the same VQS algorithm for the Green's function at zero temperature by replacing 
$\ket{\mr{G}}$ with $\ket{\Phi(\beta)}$.
It will bring out the variational quantum circuit on the original system $U_s(\vec{\theta})$ satisfying
\begin{equation}
 \begin{aligned}
&\bra{\Phi(\beta)} U_s(\vec{\theta})^\dag P_i^{(s)} U_s(\vec{\theta}) P_j^{(s)} \ket{\Phi(\beta)} \\
&\approx \bra{\Phi(\beta)} e^{iHt} P_i^{(s)} e^{-iHt} P_j^{(s)} \ket{\Phi(\beta)} \\
&= \frac{1}{Z(\beta)} \sum_n e^{-\beta E_n} \bra{E_n} e^{iHt} P_i^{(s)} e^{-iHt} P_j^{(s)} \ket{E_n},
 \end{aligned}
\end{equation}
where superscript of the Pauli operator $P_i^{(s)}$ implies that it only acts on the system $s$.
In Fig. \ref{gibbscircdynamics}, we show the ansatz quantum circuit to construct the unitary gate $U_s(\vec{\theta}(t))$.
By substiting the above quantity into Eq.~(\ref{Eq: finiteGF}), we can evaluate (each term of) $G_k(t;\beta)$.

We finally remark on another way to evaluate Eq.~\eqref{Eq: finiteGF} based on the VQS algorithm.
It is possible to obtain several approximate eigenenergies $\{\td{E}_n\}_{n=1}^K$ and eigenstates $\{\ket{\td{E}_n}\}_{n=1}^K$ of the system by the SSVQE or MCVQE algorithm, and to perform the VQS algorithm in Sec.~\ref{sec:VQS} for each obtained eigenstate $\ket{\td{E}_n}$.
The result approximates $G^R _{kn}(t)$ in Eq.~(\ref{Eq: finiteGF}), so substituting it as well as $\tilde{E}_n$ into Eq.~(\ref{Eq: finiteGF}) will give the Green's function at finite temperature (with truncating the summation up to $n=K$).
This type of the algorithm dose not need the environment qubits, but the number of energy eigenstates $K$ to evaluate the Green's function with fixed accuracy would exponentially increase as the size of the system and the inverse temperature $\beta$.

\subsection{Computing Green's function at finite temperature with excited-states search method}
Extension of the algorithm in Sec.~\ref{sec:SSVQE} to the finite temperature Green's function is rather simple.
Since we already introduce a way to evaluate the quantity $|\braket{\td{E}_n|c_{k,\uparrow}|{\td{E}_m}}|^2$, putting them into Eq.~\eqref{Eq:Lehmann_finite} gives the spectral function at finite temperature with truncating the summation up to $n=K$, where $K$ is the number of eigenstates obtained by the excited-states search algorithms.
Again, this method also has a potential problem of the exponentially increasing number of eigenstates required to compute the Green's function with fixed accuracy.

\section{Discussion and Conclusion \label{sec:Conclusion}}
In this paper, we proposed two methods for calculating the Green's function compatible with NISQ hardwares.
One of the proposed methods uses the conventional VQE to prepare a ground state, and directly calculate the real time retarded Green's function for the obtained ground state by the VQS algorithm. We introduced a method for constructing a variational quantum circuit which acts as the time evolution operator for multiple initial states simultaneously, and makes the quantum circuit for computation of the Green's function significantly shallower.
Note that this method can be straightforwardly applied to evaluation of the linear response function~\cite{Kubo1957} expressed as
\begin{equation}
\begin{aligned}
\phi_{BA}(t-t')&=i \braket{B(t-t'), A(0)}_0, \\
B(t-t')& = e^{iH (t-t')} B  e^{-iH (t-t')}
\end{aligned}
\end{equation}
where $A$ is an observable coupled to external field, for example, magnetic moment or charge density, $B$ an observable to be measured. 
The other proposed method evaluates the transition amplitude of the fermion operators between energy eigenstates of the system by exploiting either the SSVQE or MCVQE method, and computes the spectral function and the Green's function with the use of the Lehmann representation.

In numerical simulation for the Green's function at zero temperature, both methods successfully reproduced the spectral function of the 2-site Fermi-Hubbard model.
We here discuss possible causes to hinder or deteriorate the performance of the methods for general large systems.
For the first method by the VQS algorithm, the choice of the ansatz for the real time evolution is crucial: once the variational quantum state is out of the correct trajectory in the Hilbert space, it is rare that the state returns to it.
Therefore the ansatz has to be chosen carefully so that the simulated quantum state has remained in a correct trajectory.
As for the second method by the excited-states search methods, it is in general unclear how many excited states are required to reach the desired precision of the Green's function.
It will possibly grow exponentially as the inverse temperature $\beta$ and the size of the system $N$.
We leave the investigation of these problems and further comparison of two methods proposed in this study to future work. 

Since the Green's function is fundamental to study the nature of quantum systems, we believe our study will extend the possibility to utilize near-term quantum computers in condensed matter physics, quantum chemistry and materials science.

\vspace{5mm}

\textbf{Note added:}
Recently, a paper discussing calculation of the Green's function with NISQ devices appeared~\cite{rungger2019dynamical}.
The method used in that paper is based on the Lehmann representation of the Green's function, and the authors calculated excited states of the system by performing the VQE with penalty terms which are proportional to the overlaps between previously-found low excited states~\cite{higgott2019variational, jones2019variational}.

\begin{acknowledgements}
A part of the numerical simulations in this work were done on Microsoft Azure Virtual Machines provided through the program Microsoft for Startups.
YON and SE acknowledge valuable discussion with Kosuke Mitarai, Xiao Yuan, Sam McArdle, Zhenyu Cai and Takamichi Umetsu. 
IK was supported by Qunasys Inc.
This work was also supported by MEXT Q-LEAP JPMXS0118068682.
\end{acknowledgements}

\bibliography{bibliography}

\appendix

\widetext

\section{ Resource estimation of the algorithm based on the variational quantum simulation 
\label{appendix:resource}}

In this section, we discuss the number of distinct runs of quantum circuits required to implement our algorithm based on the VQS introduced in Sec.~\ref{sec:VQS}.
We firstly show that the error of the Green's function can be upper bounded by a product of the operator norm of observables involved in the Green's function and the trace distance between the ideal time-evolved state and the state computed by the VQS.
Next, based on the argument in Ref.~\cite{li2017efficient}, we discuss the errors of the real time Green's function and clarify two sources of error: algorithmic error and implementation error.
Finally, we estimate the number of total distinct runs of quantum circuits needed to achieve a certain accuracy in computing the Green's function in the frequency domain, i.e., the spectral function.

\subsection{Relationship between the error of the Green's function and the trace distance}

\subsubsection{Difference of expectation values of an observable between two states}
First, we prove the difference of expectation values of an observable $M$ for two distinct quantum states (density matrices) $\rho$ and $\sigma$ can be upper bounded by the product of the operator norm of $M$ and the trace distance of the two states.
It follows that
\begin{equation}
\begin{aligned}
\mathrm{Tr}(M(\rho-\sigma))&= \sum_k m_k \mathrm{Tr}( E_k (\rho-\sigma) ) \\
& \leq \sum_k |m_k| | \mathrm{Tr}( E_k (\rho-\sigma) )|  \\
& \leq \sum_k \|M \| \mathrm{Tr}(E_k |\rho-\sigma |) = 2 \|M \| D(\rho, \sigma),
\label{greenupper}
\end{aligned}
\end{equation}
where $\|M \|$ is an operator norm of $M$ (the largest singular value of $M$), and $M=\sum_k m_k E_k$ is the spectral decomposition of $M$ ($m_k$ is an eigenvalue of $M$ and $E_k$ is the corresponding projector).
We used $|\mathrm{Tr}(E_k (\rho-\sigma))| \leq \mathrm{Tr}(E_k |\rho-\sigma|$ \cite{nielsen2002quantum} and $\sum_k E_k =I$. 

\subsubsection{Error of the Green's function as the trace distance of the ideal state and the trial state} 
Next, we show the trace distance between the exact state after time evolution and the state obtained by the VQS gives the upper bound of the error of the calculated Green's function in our algorithm.
For simplicity, we only consider $t>0$.

Let us introduce two wavefunctions as
\begin{align}
\ket{\Psi(\vec{\theta}(t))} &= \frac{1}{\sqrt{2}} \left(\ket{0}_a \otimes U(\vec{\theta}(t)) \ket{\mathrm{G}}_s+\ket{1}_a\otimes U(\vec{\theta}(t)) P_j \ket{\mathrm{G}}_s \right), \\
\ket{\Phi(t)} &= \frac{1}{\sqrt{2}} \left(\ket{0}_a \otimes e^{-iHt} \ket{\mathrm{G}}_s+\ket{1}_a\otimes e^{-iHt} P_j \ket{\mathrm{G}}_s \right),
\label{eq:TraceDistBasic}
\end{align}
where a subscript $a$ denotes an ancilla qubit and $\ket{\mr{G}}_s$ is the ground state of the system.

In our algorithm based on the VQS, the Green's function is calculated based on the following decomposition (Eq.~\eqref{Eq:decomposedGF} in the main text),
\begin{align}
G_k^R(t) = \sum_{i,j} \lambda_i^{(k)} \lambda_j^{(k)*} \bra{\mathrm{G}}e^{-iHt} P_i e^{-iHt}P_j \ket{\mathrm{G}} \equiv \sum_{i,j} \lambda_i^{(k)} \lambda_j^{(k)*} G_{ij}^{(0)}(t).
\end{align}
The quantity $G_{ij}^{(0)}(t)$ is approximated by 
\begin{align}
G_{i,j}(\vec{\theta}(t)) \equiv  \bra{\Psi(\vec{\theta}(t))} X_a P_i \ket{\Psi(\vec{\theta}(t))}+i \bra{\Psi(\vec{\theta}(t))} Y_a P_i \ket{\Psi(\vec{\theta}(t))},
\label{eq:decomp G_ij}
\end{align}
where the measurement of $X_a P_i$ and $Y_a P_i$ correspond to the cases we set $\phi=0, \pi/2$ in Fig.~\ref{fig:efficient circuit} in the main text, respectively.
By using Eq.~\eqref{greenupper}, the error of this approximation is upper bounded as
\begin{equation}
\begin{aligned}
\epsilon_{i,j} \equiv \left| \tilde{G}_{i,j}(\vec{\theta}(t)) - G_{i,j}^{(0)}(t) \right|\lesssim 2\sqrt{2} D(\ket{\Psi(\vec{\theta}(t))},\ket{\Phi(t)})+\frac{\sqrt{2}}{\sqrt{N_m}},
\end{aligned}
\end{equation}
where $\tilde{G}_{i,j}(\vec{\theta}(t))$ is an experimentally obtained value of ${G}_{i,j}(\vec{\theta}(t))$ according to Eq.~\eqref{eq:decomp G_ij} and we denote the trace distance between two pure states $\ket{\varphi}\bra{\varphi}$ and $\ket{\varphi'}\bra{\varphi'}$ as $D(\ket{\varphi},\ket{\varphi'})$.
The second term in the right hand side indicates the shot noise, where $N_m$ is the number of measurements for $X_a P_i$ and $Y_a P_i$.
Finally, we can describe the error of the Green's function as follows:
\begin{equation}
\begin{aligned}
\epsilon^{R} \equiv \sum_{i,j} |\lambda_i^{(k)}| |\lambda_j^{(k)}| \epsilon_{i,j} &\leq \sum_{i,j} |\lambda_i^{(k)}| |\lambda_j^{(k)}| \bigg( 2\sqrt{2} D(\ket{\Psi(\vec{\theta})},\ket{\Phi(t)}_0)+\frac{\sqrt{2}}{\sqrt{N_m}}\bigg) \\
&= \alpha \bigg( 2D(\ket{\Psi(\vec{\theta}(t))},\ket{\Phi(t)}_0)+\frac{1}{\sqrt{N_m}}\bigg),
\label{eq:GF_bound}
\end{aligned}
\end{equation}
where $\alpha=\sqrt{2}\sum_{i,j} |\lambda_i^{(k)}| |\lambda_j^{(k)}|$.

\subsection{Algorithmic and implementation errors}
We now analyze sources of errors of the real time Green's function in this subsection and clarify two of main contributions of them: algorithmic and implementation errors.
According to Ref.~\cite{li2017efficient}, we can upper bound the trace distance between the ideal time-evolved state and the simulated state by the VQS as
\begin{equation}
\begin{aligned}
D\left(\ket{\Psi(\vec{\theta}(T))},\ket{\Phi(T)}_0\right) \leq D\left(\ket{\Psi(\vec{\theta}(0))},\ket{\Phi(0)}_0 \right) + \sum_{n=1}^{N_{\mr{step}}} D\left(U\ket{\Psi(\vec{\theta}((n-1)\delta t))}, \ket{\Psi(\vec{\theta}(n \delta t))}\right),
\end{aligned}
\end{equation}
where $T$ is the time to be simulated, $U$ is the exact time evolution operator of small time step $\delta t$ described as $U=\mathrm{exp}(-iH \delta t)$, and $N_{\mathrm{step}}=T/\delta t$ is the number of time steps.
The first term corresponds to the state preparation error of the ground state via the variational quantum eigensolver (VQE), which we denote as $\epsilon_s=D\left(\ket{\Psi(\vec{\theta}(0))},\ket{\Phi(0)}_0 \right)$.
The second term, which accumulates an error for each time step, can be decomposed into two types of errors: algorithmic and implementation errors.
Algorithmic errors contains error due to imperfection of the ansatz approximating the trial state and one due to a finite time step.
Implementation errors are caused by physical errors stemming from the noisy nature of near-term quantum computers and the shot noise.
We assume physical errors can be ignored because it can be suppressed by using quantum error mitigation techniques~\cite{temme2017error,endo2018practical,mcardle2019error,song2019quantum,kandala2019error,zhang2020error}, i.e., we consider only the shot noise as the implementation error.
By using the triangle inequality, it follows that
\begin{equation}
D\left(U\ket{\Psi(\vec{\theta}((n-1)\delta t))}, \ket{\Psi(\vec{\theta}(n \delta t))}\right)  \leq D\left(U\ket{\Psi(\vec{\theta}((n-1)\delta t))}, \ket{\Psi^{(0)}(\vec{\theta}(n \delta t))}\right) + D\left(\ket{\Psi(\vec{\theta}(n \delta t))}, \ket{\Psi^{(0)}(\vec{\theta}(n \delta t))}\right),
\end{equation}
 where $\ket{\Psi^{(0)}}$ denotes the state without implementation errors.
 The first term in the right hand side corresponds to the algorithmic error and the second term does to the implementation error for each time step.
 We denote these errors $\delta \epsilon_A(n)$ and $\delta \epsilon_I(n)$, respectively.

\subsubsection{Algorithmic error}
The algorithmic error can be written as~\cite{li2017efficient} 
\begin{equation}
\delta \epsilon_A(n) \equiv D\left(U\ket{\Psi(\vec{\theta}(n-1)\delta t))}, \ket{\Psi^{(0)}(\vec{\theta}(n\delta t))}\right) = \sqrt{\Delta ^{(2)}_n \delta t^2 +\Delta ^{(3)}_n \delta t^3 + O(\delta t^4)}, 
\end{equation}
where 
\begin{equation}
 \begin{aligned}
 \Delta^{(2)}_n &= \braket{\delta \Psi(\vec{\theta}(n \delta t))|\delta \Psi(\vec{\theta}(n \delta t))} - \left|\braket{\delta \Psi(\vec{\theta}(n \delta t))| \Psi(\vec{\theta}((n-1)\delta t))} \right|^2, \\
 \ket{\delta \Psi(\vec{\theta}(n \delta t))}&=-iH \ket{ \Psi(\vec{\theta}((n-1)\delta t))}-\sum_k \dot{\theta}_k^{(0)}\frac{\partial \ket{\Psi(\vec{\theta}((n-1)\delta t))}}{\partial \theta_k}, \\
 \Delta^{(3)}_n &= \|H\| \|H^2\| + \frac{1}{3}\|H^3\| + \|\frac{d\tilde{R}}{dt}\| \|\frac{d^2\tilde{R}}{dt^2}\| + \frac{1}{3}\|\frac{d^3\tilde{R}}{dt^3}\| \\
 & \quad + \|H^2\|\left(\|\frac{d\tilde{R}}{dt}\|^2 + \|\frac{d^2\tilde{R}}{dt^2}\|\right) + \left( \|H\| + \|H^2\| \right) \|\frac{d\tilde{R}}{dt}\|,
 \end{aligned}
\end{equation}
the matrix norm $\|\ldots\|$is induced by the vector norm, the operator $d/dt$ is defined as $d/dt = \dot{\vec{\theta}}(t) \cdot \partial/\partial \vec{\theta}$, and $\tilde{R} \equiv = I_a \otimes U(\vec{\theta}(t))$ is the unitary circuit consisting of the ansatz (Eq.~\eqref{eq:Ansatz for geneal VQS}). 
The term proportional to $\Delta^{(2)}_n$ stems from limited representative capability of the ansatz state $\ket{\Psi^{(0)}(\vec{\theta})}$ approximating the true state.
The term proportional to $\Delta^{(3)}_n$ is, on the other hand, due to a finite time step $\delta t$. 
Therefore, the total accumulation of algorithmic errors from $t=0$ to $t=T$ is 
\begin{align}
\epsilon_A= \sum_{n=1}^{N_{\mathrm{step}}} \delta \epsilon_A(n) \lesssim \sqrt{\Delta_{\mathrm{max}}^{(2)}}T +\sqrt{\Delta_{\mathrm{max}}^{(3)} \delta t},
\label{eq:AlgoError}
\end{align}
where $\Delta^{(2)}_{\mr{max}} = \max_n \Delta^{(2)}_n$ and $\Delta^{(3)}_{\mr{max}} = \max_n \Delta^{(3)}_n$.
The first term in the right hand side does not depend on the time step $\delta t$ because this is the error caused by imperfections of the ansatz to represent the quantum state after the time evolution, whereas the second term can be suppressed by taking the small time step.

Here, we also show another important property of our algorithm based on the VQS.
By denoting $\ket{\Psi(\vec{\theta}(t))}$ as $\ket{\Psi(\vec{\theta}(t))} = \frac{1}{\sqrt{2}}\left(\ket{0}_a \otimes \ket{\psi_1(\vec{\theta}(t))}+\ket{1}_a \otimes \ket{\psi_2(\vec{\theta}(t))}\right)$ for simplicity, we have
\begin{equation}
\begin{aligned}
\Delta^{(2)}_n &=\frac{1}{2}\left(\braket{\delta \psi_1(\vec{\theta}(n \delta t))| \delta \psi_1(\vec{\theta}(n \delta t))}+\braket{\delta \psi_2(\vec{\theta}(n \delta t))| \delta \psi_2(\vec{\theta}(n \delta t))}\right) \\
&-\frac{1}{4}\left( \left|\braket{\delta \psi_1(\vec{\theta}(n \delta t))| \psi_1(\vec{\theta}(n \delta t))}\right|^2 + \left|\braket{\delta \psi_2(\vec{\theta}(n \delta t))| \psi_2(\vec{\theta}(n \delta t))}\right|^2 \right)\\
&-\frac{1}{2} \mathrm{Re}\left(\braket{\delta \psi_1(\vec{\theta}(n \delta t))| \psi_1(\vec{\theta}(n \delta t))}\braket{\delta \psi_2(\vec{\theta}(n \delta t))| \psi_2(\vec{\theta}(n \delta t))}\right) \\
&=\frac{1}{2}(\Delta^{(2)}_{n, \psi_1}+\Delta^{(2)}_{n, \psi_2}) + \frac{1}{4}\left|\braket{\delta \psi_1(\vec{\theta}(n \delta t))|\psi_1(\vec{\theta}(n \delta t))}-\braket{\delta \psi_2(\vec{\theta}(n \delta t))|\psi_2(\vec{\theta}(n \delta t))}\right|^2, 
\end{aligned}
\end{equation}
where 
\begin{equation}
\begin{aligned}
 \Delta^{(2)}_{n, \psi_i} &= \braket{\delta \psi_i(\vec{\theta}(n \delta t))|\delta \psi_i(\vec{\theta}(n \delta t))} - \left|\braket{\delta \psi_i(\vec{\theta}(n \delta t))| \psi_i(\vec{\theta}((n-1)\delta t))}\right|^2, \\
 \ket{\delta \psi_i(\vec{\theta}(n \delta t))}&=-iH \ket{ \psi_i(\vec{\theta}((n-1)\delta t))}-\sum_k \dot{\theta}_k^{(0)}\frac{\partial \ket{\psi_i(\vec{\theta}((n-1)\delta t))}}{\partial \theta_k}.
 \end{aligned}
 \end{equation}
 From the expression above, one can see that $\Delta^{(2)}_n$ for $\ket{\Psi}$ is not just a mean of $\Delta^{(2)}_n$s for $\ket{\psi_1}$ and $\ket{\psi_2}$ but also includes the term $\delta_{12} = \frac{1}{4}|\braket{\delta \psi_1|\psi_1}-\braket{\delta \psi_2|\psi_2}|^2$.
 This term, $\delta_{12}$, indicates that the algorithmic error due to the insufficient ansatz increases when we  want to find a unitary operator to evolve two input states simultaneously, compared with finding different unitaries for each state.
The term $\delta_{12}$ clarifies the drawback of our algorithm to find a unitary which simultaneously approximates the time evolution operator for multiple states.

However, as seen in numerical simulation in Sec.~\ref{sec:Numerics}, we successfully reproduced the real time Green's functions of the two site fermi-Hubbard model, which implies that the effect of $\delta_{12}$ can be made small.
In general, the cost for employing different unitaries for multiple states and the circuit like in Fig.~\ref{fig:naive circuit}
is much larger than employing a single unitary and the circuit like in Fig.~\ref{fig:efficient circuit}.

\subsubsection{Implementation error}
The implementation error is caused by the shot noise and give errors in the solution of Eq.~\eqref{Eq:MthetaV} in the main text,
\begin{align}
M \dot{\vec{\theta}}=\vec{V}, 
\label{appeq:MthetaV}
\end{align}
where 
\begin{align}
M_{i,j}& = \left. \mathrm{Re}\bigg(\frac{\partial \bra{\Psi(\vec{\theta})}}{\partial \theta_i} \frac{\partial \ket{\Psi(\vec{\theta})}}{\partial \theta_j} \bigg) \right|_{\vec{\theta}=\vec{\theta}(n\delta t)}, \\
V_j&= \left. \mathrm{Im}\bigg(\bra{\Psi(\vec{\theta})}H \frac{\partial \ket{\Psi(\vec{\theta})}}{\partial \theta_j} \bigg) \right|_{\vec{\theta}=\vec{\theta}(n\delta t)}.
\end{align}
At each time step, the matrix $M$ and the vector $V$ are obtained by measuring outputs of appropriate quantum circuits~\cite{li2017efficient, PhysRevResearch.1.013006}, which include the shot noise.
We denote error-free values of the matrix $M$ and the vector $\vec{V}$ as $M_0$ and $\vec{V}_0$, and observed values of them as $M_0 + \delta M$ and $\vec{V}_0 + \delta \vec{V}$, respectively.
The solution of Eq.~\eqref{appeq:MthetaV} calculated by $M_0$ and $\vec{V}_0$ is denoted by $\dot{\vec{\theta}}_0$ and the one calculated by $M_0+\delta M$ and $\vec{V}_0 + \delta \vec{V}$ as $\dot{\vec{\theta}}_0 + \delta \dot{\vec{\theta}}$.
In the first order of $\delta M_0$ and $\delta \vec{V}_0$, it follows 
\begin{equation}
\delta \dot{\vec{\theta}} \approx M_0^{-1} \delta \vec{V} -M_0^{-2} \delta M \vec{V}_0,
\end{equation}
and we thus have 
\begin{align}
\|\delta \dot{\vec{\theta}}\| \leq \| M_0^{-1}\| \|\delta \vec{V}\| + \|M_0^{-1}\|^2 \|V_0\| \|\delta M\|.
\end{align}
Since we consider only the shot noise, $\|\delta \vec{V}\|$ and $\|\delta M\|$ can be described as
\begin{align}
\|\delta M\| &\approx \frac{\Delta_M}{\sqrt{N_r}}, \\
\|\delta \vec{V}\| & \approx  \frac{\Delta_{\vec{V}}}{\sqrt{N_r}}, 
\end{align}
where $N_r$ is the number of shots to evaluate $M$ and $\vec{V}$.
When we denote the derivative of the state $\ket{\Psi(\vec{\theta})}$ as
\begin{align}
\frac{\partial \ket{\Psi(\vec{\theta})}}{\partial \theta_k}=\sum_{i=1}^{N_{D,k}} g_{k,i} I_a \otimes \left(U_N\cdots U_{k+1} P_{k,i} U_{k-1}\cdots U_1\right) \ket{\Psi(\vec{\theta})},
\end{align}
where $g_{k,i}$ is a complex number and $P_{k,i}$ is a Pauli operator (recalling that we have assumed the ansatz has the form of Eq.~\eqref{eq:ansatz} in the main text),
$\Delta_M$ can be written as
\begin{equation}
\Delta_M = 2 \sqrt{ \sum_{k,q} \left(\sum_{i,j} |g_{k,i}^* g_{q,j}|^2\right)}.
\end{equation} 
Similarly, $\Delta_{\vec{V}}$ is written as
\begin{equation}
\Delta_{\vec{V}}=2\sqrt{ \sum_k \left(\sum_{i,j}|g_{k,i}^* h_j|^2\right) },
\end{equation} 
where the Hamiltonian is written as $H=\sum_{j} h_j P_j$ with $P_j$ being a Pauli operator and $h_j$ a coefficient.

In Ref.~\cite{li2017efficient}, the implementation error is shown to be approximated as 
\begin{equation}
\begin{aligned}
\delta \epsilon_I(n) &\equiv  D\left(\ket{\Psi(\vec{\theta}(n \delta t))}, \ket{\Psi^{(0)}(\vec{\theta}(n \delta t))}\right) \\
&= \sqrt{\delta \dot{\vec{\theta}}^T B \delta \dot{\vec{\theta}} \delta t^2 +O(\delta t^3)} \\
&\lesssim \sqrt{\|B \|} \|\delta \dot{\vec{\theta}} \| \delta t  \\
&\lesssim \sqrt{\|B\|} \frac{\Delta_I}{\sqrt{N_r}} \delta t,
\end{aligned}
\end{equation}
where $\Delta_I=\|M_0^{-1} \| \Delta_{\vec{V}}+\|M_0^{-1}\|^2 \| V_0\| \Delta_M $ and 
\begin{align*}
B_{i,j} &= \frac{\partial \bra{\Psi(\vec{\theta}((n-1)\delta t))}}{\partial \theta_i} \frac{\partial \ket{\Psi(\vec{\theta}((n-1) \delta t))}}{\partial \theta_j} \\
 & \quad - \frac{\partial \bra{\Psi(\vec{\theta}((n-1) \delta t))}}{\partial \theta_i} \ket{\Psi(\vec{\theta}((n-1)\delta t))}\bra{\Psi(\vec{\theta}((n-1) \delta t))}  \frac{\partial \ket{\Psi(\vec{\theta}((n-1) \delta t))}}{\partial \theta_j}.
\end{align*}
Finally, the total implementation error will be
\begin{align}
\epsilon_I \equiv \sum_{n=1}^{N_\mr{step}} \delta \epsilon_I(n) \leq \sqrt{\|B\|_{\mr{max}}} \frac{\Delta_I^{(\mathrm{max})}}{\sqrt{N_r}}T,
\label{eq:ImpleError}
\end{align}
where the superscipt and subscript ``max" on $\|B\|$ and $\Delta_I$  represents the maximum value for $n=1,\ldots,N_\mr{step}$.

\subsection{Resource estimation}
\subsubsection{Evaluation of the error of the real time Green's function and the resource estimation}
From Eqs.\eqref{eq:GF_bound}\eqref{eq:AlgoError}\eqref{eq:ImpleError}, the error of the computed Green's function by our algorithm based on the VQS reads as
\begin{equation}
\begin{aligned}
\epsilon^R &\leq \alpha \left( 2(\epsilon_s+\epsilon_A+\epsilon_I)+\frac{1}{\sqrt{N_m}}\right) \\
&=\alpha \left( 2 \left(\epsilon_s+\sqrt{\Delta^{(2)}}T+\sqrt{\Delta^{(3)}\delta t}T +\sqrt{\|B\|} \frac{\Delta_I}{\sqrt{N_r}}T \right) + \frac{1}{\sqrt{N_m}} \right),
\end{aligned}
\label{eq:ErrorTotal}
\end{equation}
where we omitted the subscript and superscript ``max" for simplicity. The first and second terms in the right hand correspond to the state preparation error and the (part of) algorithmic error, respectively.
Both stems from imperfections of the ansatz to represent the ideal quantum state to be simulated and the analysis of them is not straightforward; e.g., it depends on details of the system (Hamiltonian) and/or the ansatz as well as the optimisation method used for the VQE.  We leave them for the future work.
On the other hand, the third, forth, and fifth terms in the right hand side can be suppressed by employing sufficiently small time step $\delta t$ and the large number of shots $N_r, N_m$ for measuring the quantum circuits.
Those terms represent the algorithmic error due to a finite time step, the implementation error, and the shot noise in evaluating the Pauli matrices involved in the Green's function, respectively.

Let us evaluate the size of time step $\delta t$ and the number of shots $N_r, N_m$ so as to bound each term in the right hand side of Eq.~\eqref{eq:ErrorTotal}.
To upper bound the third term by $\varepsilon_A$, we need the time step
\begin{equation}
\delta t= \frac{\varepsilon_A^2}{4 \alpha^2 \Delta^{(3)}T^2}. 
\label{timestep}
\end{equation}
It means that the number of time steps will be
\begin{align}
N_\mr{step}=\frac{T}{\delta t}=\frac{4\alpha^2 \Delta^{(3)} T^3}{\varepsilon_A^2}.
\end{align}
To upper bound the fourth term by $\varepsilon_I$, we need to set the number of shots to evaluate the matrix $M$ as
\begin{align}
N_r= \frac{4\alpha^2 \|B \|T^2}{\varepsilon_I^2}.
\end{align}
Similarly, if we want to upper bound the fifth term by $\varepsilon_m$, it is required
\begin{align}
N_m = \frac{\alpha^2}{\varepsilon_m^2}. 
\end{align}

Finally, let us count the number of distinct runs of the quantum circuits to obtain the Green's function and discuss its dependence on the required error bound to the Green's function. 
The number of shots (distinct runs of quantum circuit) to populate $M$ and $V$ matrix is at most $N_{\theta}^2 N_D^2 + N_{\theta} N_H N_D$, where $N_\theta$ is the number of parameters (the number of elements of $\vec{\theta}$), $N_H$ is the number of terms of the Hamiltonian, and $N_D$ is $\max_k N_{D,k}$.
Hence the number of distinct runs of the quantum circuits to obtain the Green's function will be 
\begin{align}
N_{\mathrm{tot}}&= N_\mr{step} \times N_r \times (N_{\theta}^2 N_D^2 + N_{\theta} N_H N_D) + N_k \times N_m \nonumber \\
&= \frac{16 \alpha^4 \|B\|\Delta^{(3)}T^5}{\varepsilon_A^2\varepsilon_I^2} (N_{\theta}^2 N_D^2 + N_{\theta} N_H N_D) + \frac{N_k\alpha^2}{\varepsilon_m^2},
\label{totalresource}
\end{align}
where $N_k$ is the number of the Pauli matrices in which we expand the electron creation operator $c_k$ (see Eq.~\eqref{eq:ExpandElectron} in the main text).
This is one of the main results for the resource estimation for calculating the Green's function based on the VQS, although this number is a very loose, or pessimistic estimation.

We note that further simplification may be possible when we set $\varepsilon_A=\varepsilon_I=\varepsilon_m=\varepsilon/3$, i.e., we set the total error of the Green's function to $\varepsilon$, ignoring the errors originating from imperfections of the ansatz,
\begin{align}
N_{\mathrm{tot}} &= \frac{1296 \alpha^4 \|B\|\Delta^{(3)}T^5}{\varepsilon^4} (N_{\theta}^2 N_D^2 + N_{\theta} N_H N_D) + \frac{9N_k\alpha^4}{\varepsilon^2} \nonumber \\
&\approx \frac{1296 \alpha^4 \|B\|\Delta^{(3)}T^5}{\varepsilon^4} (N_{\theta}^2 N_D^2 + N_{\theta} N_H N_D). 
\end{align}
We will use this relationship in the next subsubsection.

\subsubsection{Evaluation of the error of the Green's function in frequency domain and the resource estimation}
The Green's function in frequency domain is defined as
\begin{equation}
  \tilde{G}^R_k(\omega)= \int_{0}^{\infty} e^{i (\omega+i\eta) t} G_k^R(t) dt~ (\eta \rightarrow +0),    
\end{equation}
and its imaginary part $\mr{Im}\,\tilde{G}^R_k(\omega)$ gives the spectrum function.
Suppose the Hamiltonian of interest has the energy spectrum $\{ E_\mr{min}, \ldots, E_\mr{max}\}$, where $E_\mr{min(max)}$ is the minimal (maximum) energy. We assume all energies are positive, $E_\mr{min}>0$, without loss of generality.
According to the sampling theorem, to avoid aliasing in $\tilde{G}^R_k(\omega)$, we need to sample the real time Green's function $G_k^R(t)$ with over twice the maximum frequency contained in it.
It means that we need to have the time step 
\begin{align}
\delta t \leq \frac{1}{2 f_{\mathrm{max}}}= \frac{1}{{E}_{\mathrm{max}}/ \pi},
\end{align}
where $f_\mr{max} = E_\mr{max}/2\pi$ is the maximum frequency of the real time Green's function, in the VQS algorithm.
It is important to note that this bound is constant with regard to the required accuracy to the real time Green's function while the time step shown in Eq.~\eqref{timestep} depends on it,
so we adopt Eq.~\eqref{timestep} as the time step to be considered. 
Meanwhile, the necessary evolution time $T_0$ to obtain information of the lowest frequency component is $T_0=2\pi/E_{\mathrm{min}}$.

In our simple numerical integration scheme, the real time Green's function is sampled at $t=0, \delta t, \ldots, T_0$.
The Green's function in frequency domain is approximated as  
\begin{equation}
\begin{aligned}
\tilde{G}^R_k(\omega) &\approx \int_{0}^{T_0} e^{i (\omega+i\eta) t} G_k^R(t) dt~ (\eta \rightarrow +0) \\
&\approx \sum_{n=0}^{N_{\mathrm{step}}} G_k^R(n \delta t) e^{i \omega n \delta t} \delta t + O(\delta t) 
\end{aligned}
\end{equation}
where $N_{\mathrm{step}} = T_0/\delta t$ is the total number of time steps for simulation.
We note that the first term in the last line is $O(\delta t^0)$.
When the errors of all data $\{ G_k^R(n \delta t) \}_{n=0}^{N_\mr{step}}$ are bounded by $\tilde{\varepsilon}$, the error of the imaginary part of $\tilde{G}^R_k(\omega)$ can be upper bounded as
\begin{equation}
\delta \left( \mr{Im}[\tilde{G}_k^R(\omega)] \right)
 \lesssim N_\mr{step} \cdot  \tilde{\varepsilon} \cdot \delta t
 = \tilde{\varepsilon} \, T_0.
\end{equation}
The same argument can be made for the real part.
Therefore, if we want to set total accuracy of $\tilde{G}^{(R)}_k(\omega)$ to $\varepsilon$, i.e., we should take $\tilde{\varepsilon} = \varepsilon / T_0$.
In that case, the total number of measurements (Eq.~\eqref{totalresource}) will be 
\begin{equation}
\begin{aligned}
N_{\mathrm{tot}}&= \frac{1296 \alpha^4 \|B\|\Delta^{(3)}T_0^9}{\varepsilon^4}(N_\theta^2 N_D^2 + N_\theta N_H N_D) \\
&= \frac{(2\pi)^9 \times 1296  \alpha^4 \|B\|\Delta^{(3)}}{\varepsilon^4 E_{\mathrm{min}}^9}(N_\theta^2 N_D^2 + N_\theta N_H N_D).
\end{aligned}
\end{equation}

Conversely, when the time step $\delta t$ is fixed and take only errors related to $\delta t$ into account, the error of the Green's function in frequency domain is
\begin{equation}
\delta \left( \mr{Im}[\tilde{G}_k^R(\omega)] \right)
 \sim 2\alpha T_0^{3/2} \sqrt{\Delta^{(3)} \delta t}.
\end{equation}

\section{Details of discussion on the feasibility of our algorithms in Sec.\ref{sec:gatecount}}
\label{appendix:gatecount}

In this section, we provide details of the discussion on the feasibility of our proposed algorithms in Sec.~\ref{sec:gatecount} based on Ref.~\cite{cai2019resource}.

We take the the two-dimensional Hubbard model (Eq.~\eqref{eq:2dHubbardDef} in the main text) of $N_{\mathrm{site}}$ sites with the open boundary condition as an example.
By applying the Jordan-Wigner transformation, the Hamiltonian will have the form of $H=\sum_j h_j P_j $, where $h_j$ is real coefficient and $P_j$ is Pauli matrices.
The Hamiltonian ansatz is defined as (Eq.~\eqref{HamAnsatz_def} in the main text)
\begin{align}
 \prod_{d=1}^{n_d} \left( \prod_j \exp\left( i \theta_j^{(d)} P_j \right) \right),
\end{align}
where $n_d$ is the depth of the ansatz and $\theta_j^{(d)}$ is a parameter to be optimized in the ansatz.
In Ref.~\cite{cai2019resource}, it is shown that the number of single qubit gates and two-qubit gates to implement the single-depth ($n_d=1$) Hamiltonian ansatz of this model is
\begin{equation}
\begin{aligned}
N_{\mathrm{single}}&=4 N_{\mathrm{site}}^{\frac{3}{2}}+7  N_{\mathrm{site}}-4 \sqrt{N_{\mathrm{site}}}, \\
N_{\mathrm{two}}&=8 N_{\mathrm{site}}^{\frac{3}{2}}+ N_{\mathrm{site}}-4 \sqrt{N_{\mathrm{site}}},
\end{aligned}
\end{equation}
with assuming the Z rotation gate and the partial swap gate as elementary single and two-qubit operations.
Therefore, the total number of gates is $n_d(N_{\mathrm{single}}+N_{\mathrm{two}}) $.
By putting $n_d = 1$ and $N_\mr{site}=25$ as in the main text, we obtain $N_{\mr{single}} \approx 650, N_{\mr{two}} \approx 1000$.
Alternatively, when we set $n_d=N_\mr{site}$, the total number of gates is $N_{\mr{single}} \approx 16400, N_{\mr{two}}\approx  2600$.
In this case, the tolerable error rate for quantum error correction, $N_{\mathrm{gate}}\epsilon_{\mathrm{gate}} \lesssim 2$, gives $\epsilon_{\mathrm{gate}} = 8 \times 10^{-2} \%$ even if we consider only the two-qubit gates.
This value indicates that the further  improvement of two-qubit gate fidelity is required.

\section{Additional simulation results}
\label{additional}
\begin{figure*}[h!]
     \includegraphics[width=.3\textwidth, trim=20 0 20 20]{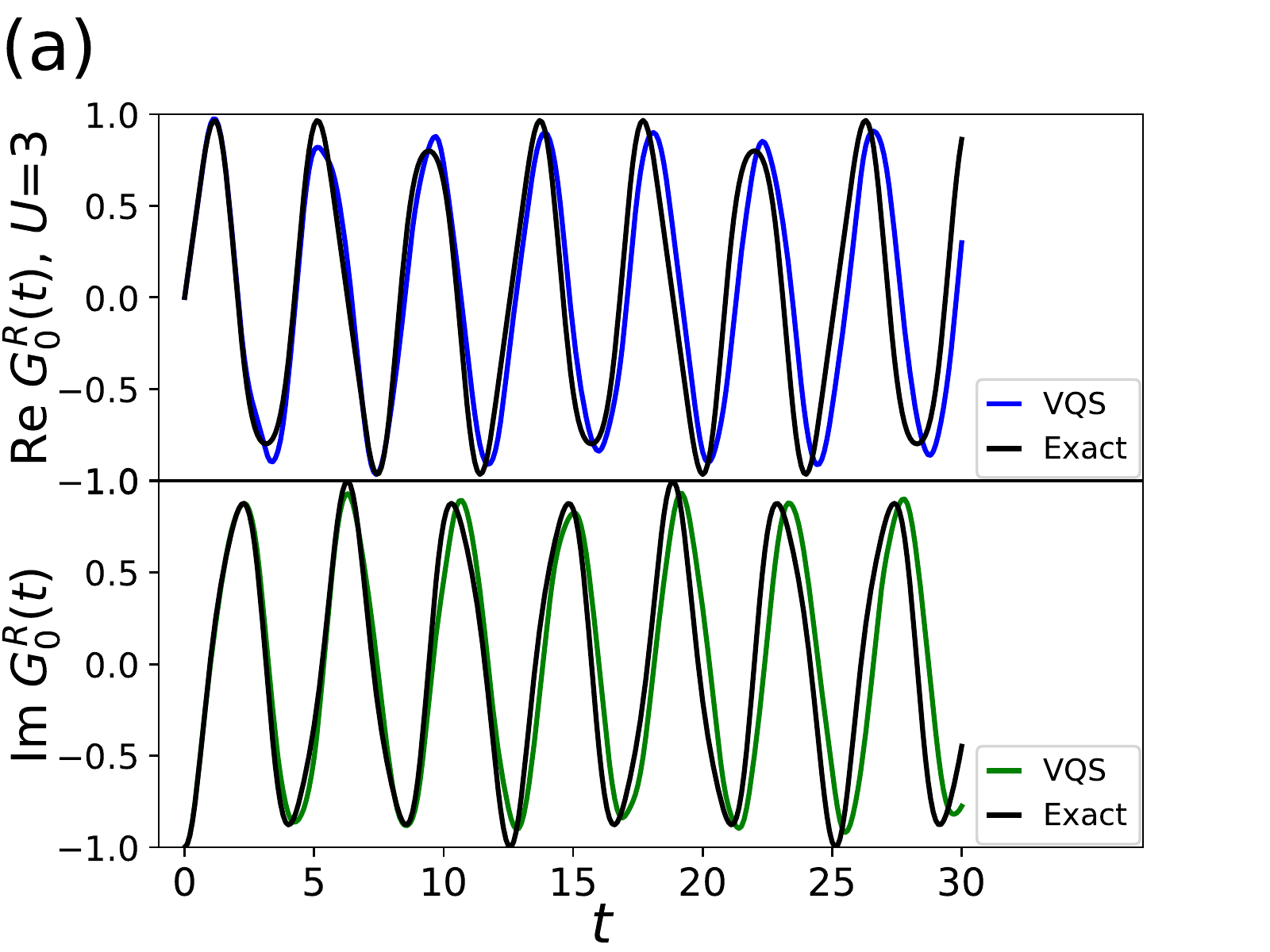}
     \includegraphics[width=.3\textwidth, trim=20 0 20 20]{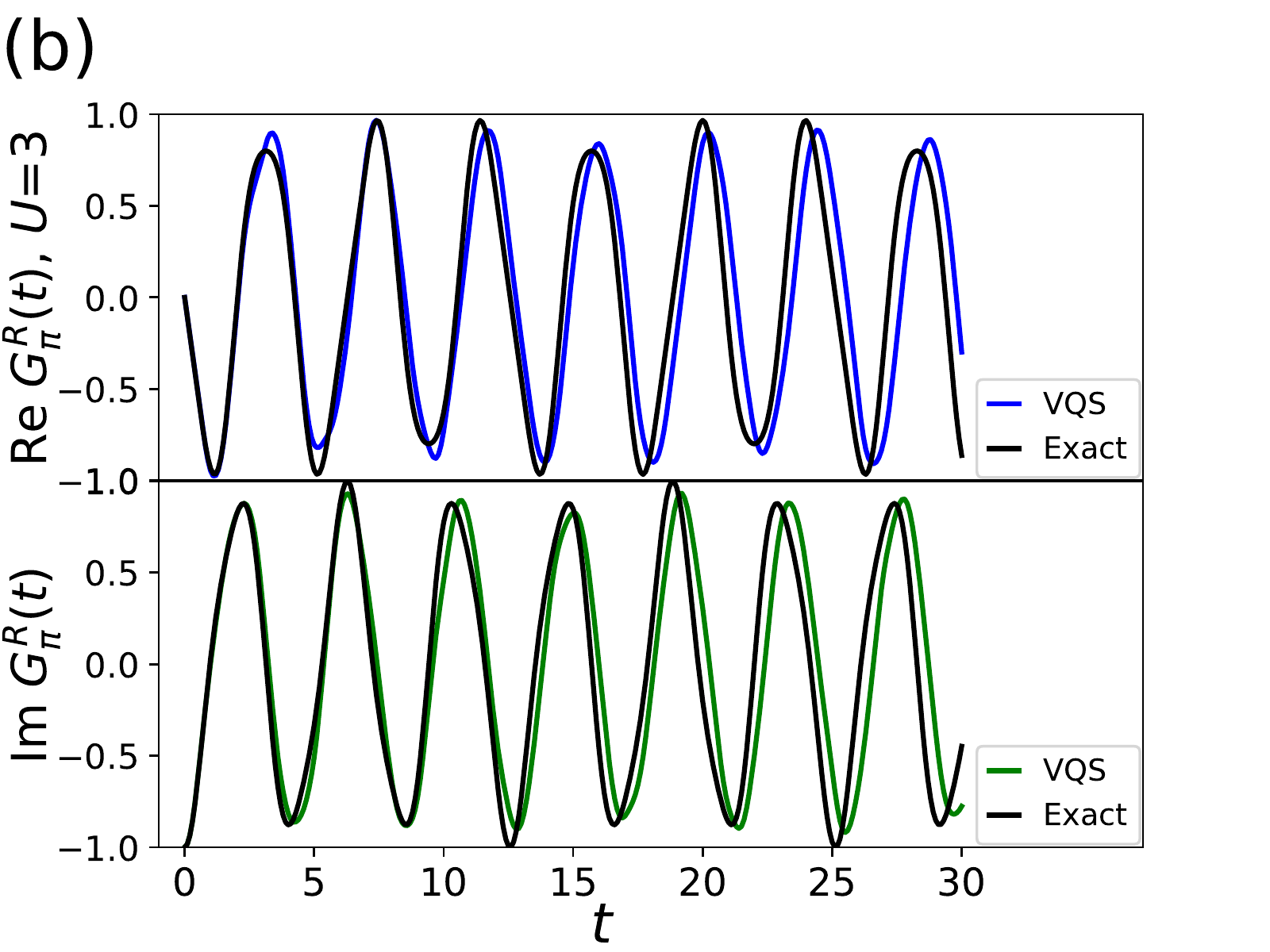}
     \includegraphics[width=.3\textwidth, trim=20 0 20 20]{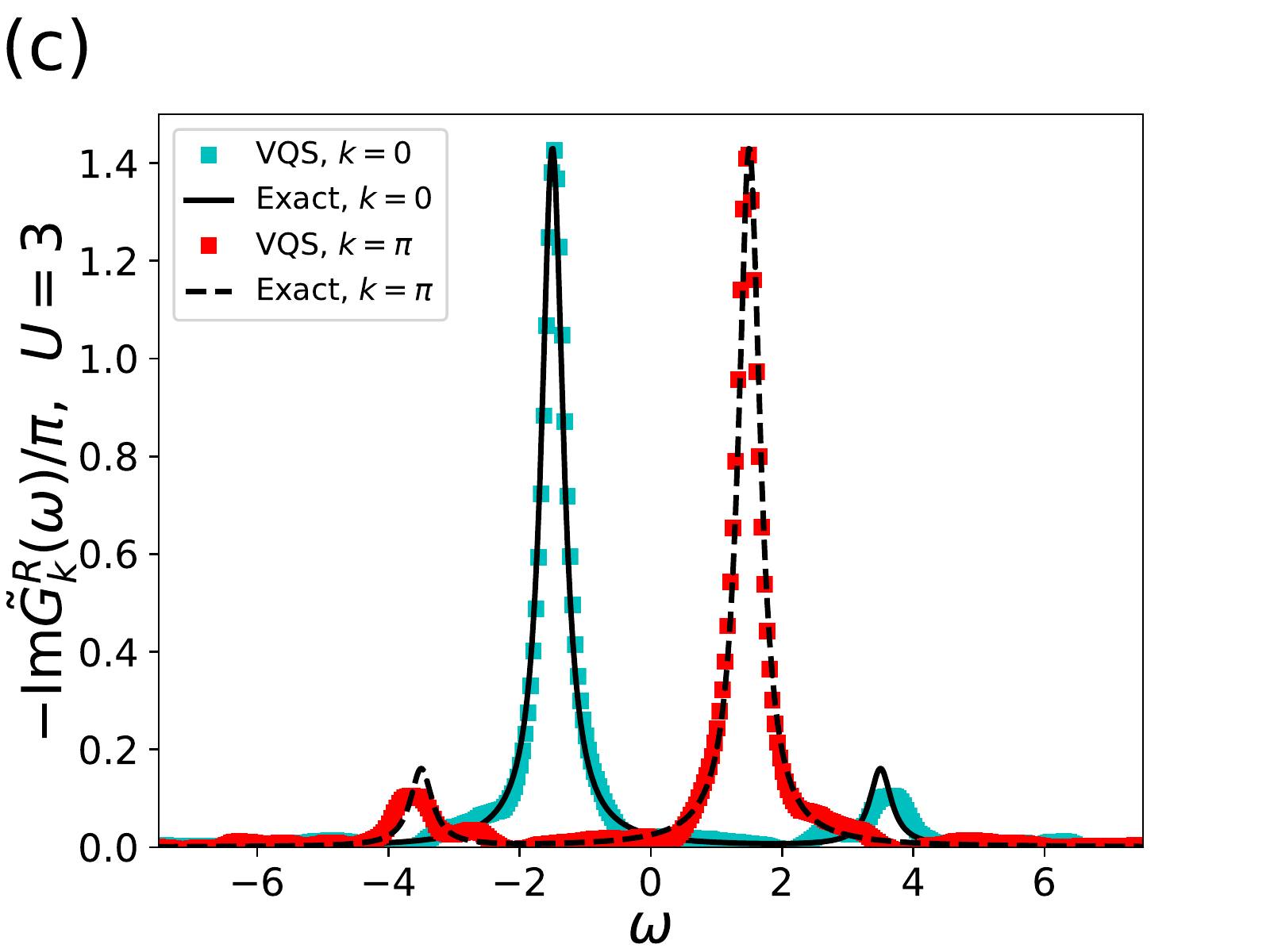} \\
     \includegraphics[width=.3\textwidth, trim=20 20 20 20]{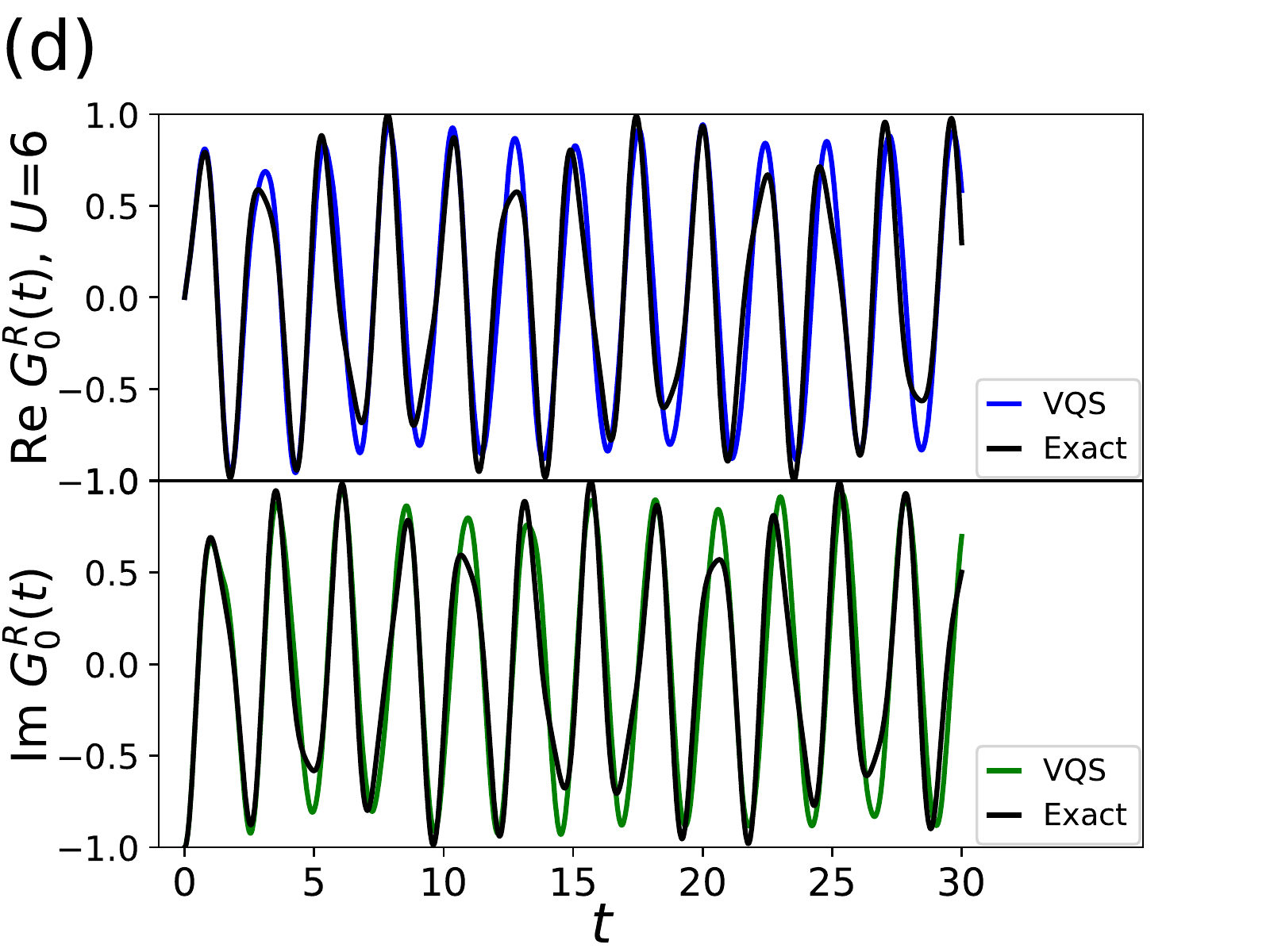}
     \includegraphics[width=.3\textwidth, trim=20 20 20 20]{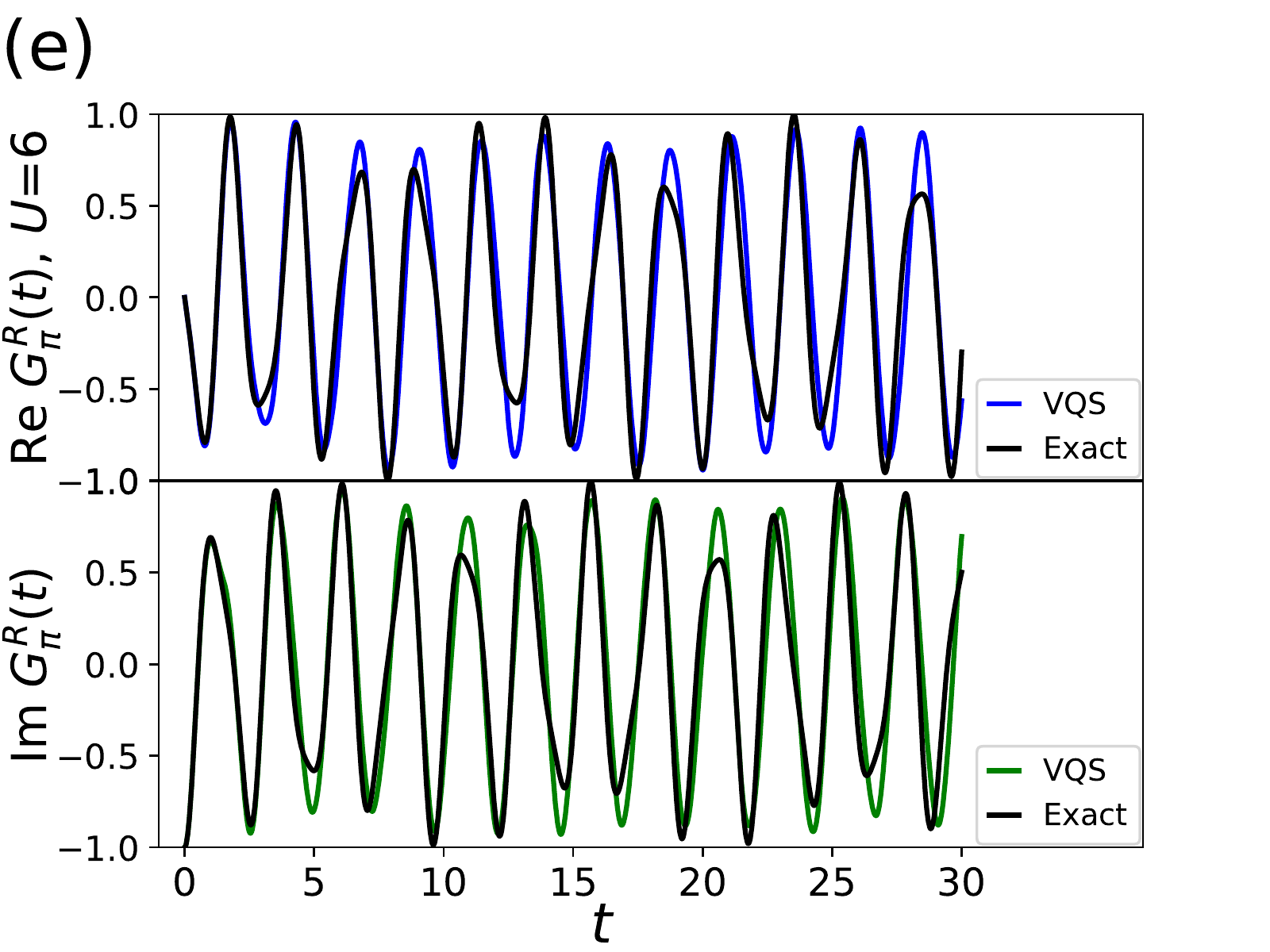}
     \includegraphics[width=.3\textwidth, trim=20 20 20 20]{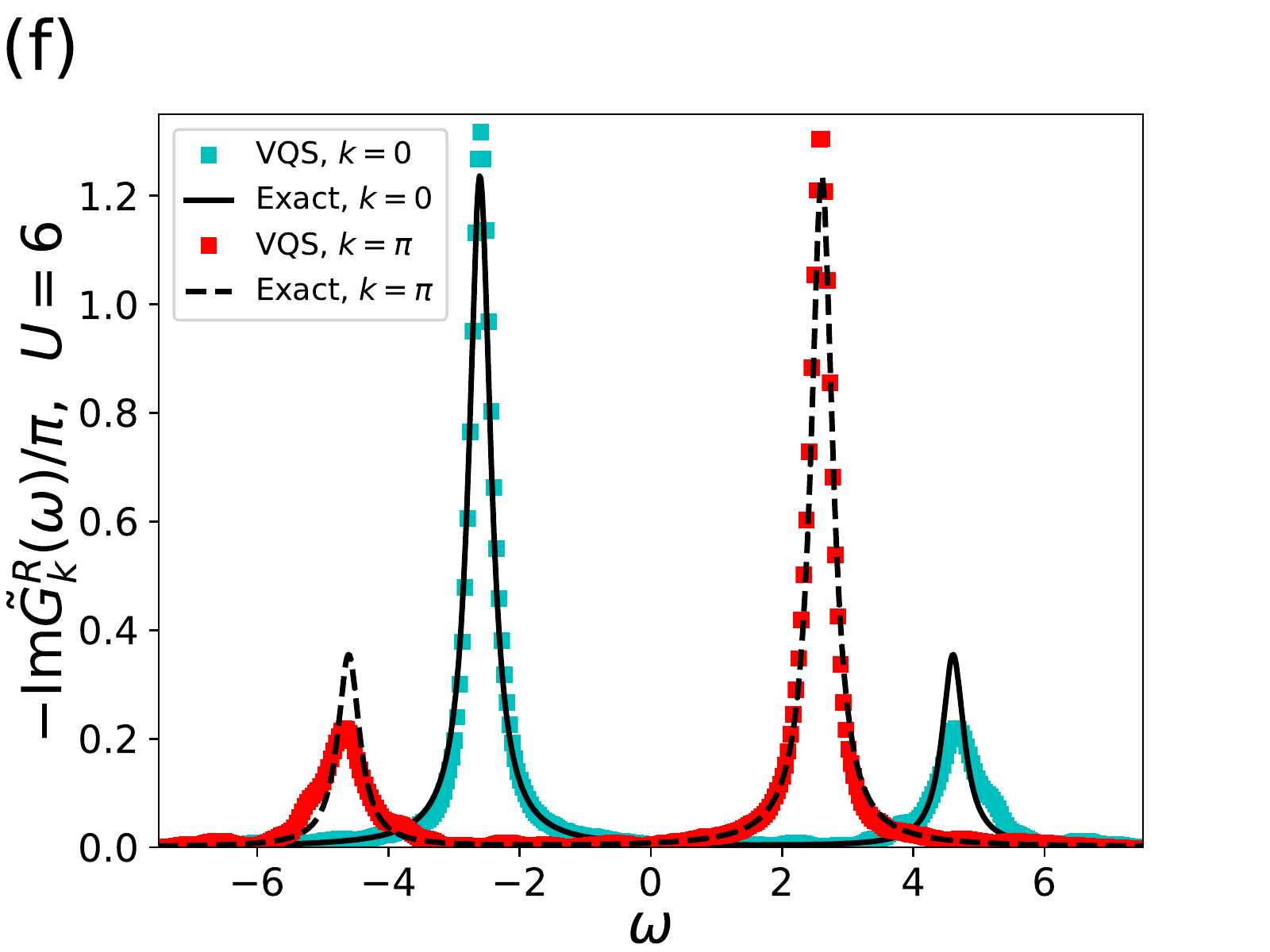}
   \caption{\label{appfig:VQS result}
   Numerical simulation of the VQS algorithm to compute the Green's function in real time $G^R_k(t)$ (a, b, d, e) and the spectral function (c, f) for $n_d=4$ and the model~\eqref{Hubbard_def} of $U=3$ (a-c) and $U=6$ (d-f). 
   The time step is taken as $dt=0.1 (0.03)$ for $U=3 (6)$.
   The exact spectral function is calculated by the exact dynamics of the Green's function in real time from $t=0$ to $t=100$ with step $dt=0.1$.
   We take $\eta = 0.2$ for the calculation of the spectral functions.}
\end{figure*}

Here, we show additional numerical results. The results for $n_d=4$ are presented in Fig.~\ref{appfig:VQS result}. We can see that the results are almost identical with the result of $n_d=8$ in Fig.~\ref{fig:VQS result} in the main text although the peaks of the spectral function are slightly smeared.

\begin{figure}[h!]
     \includegraphics[width=.45\textwidth]{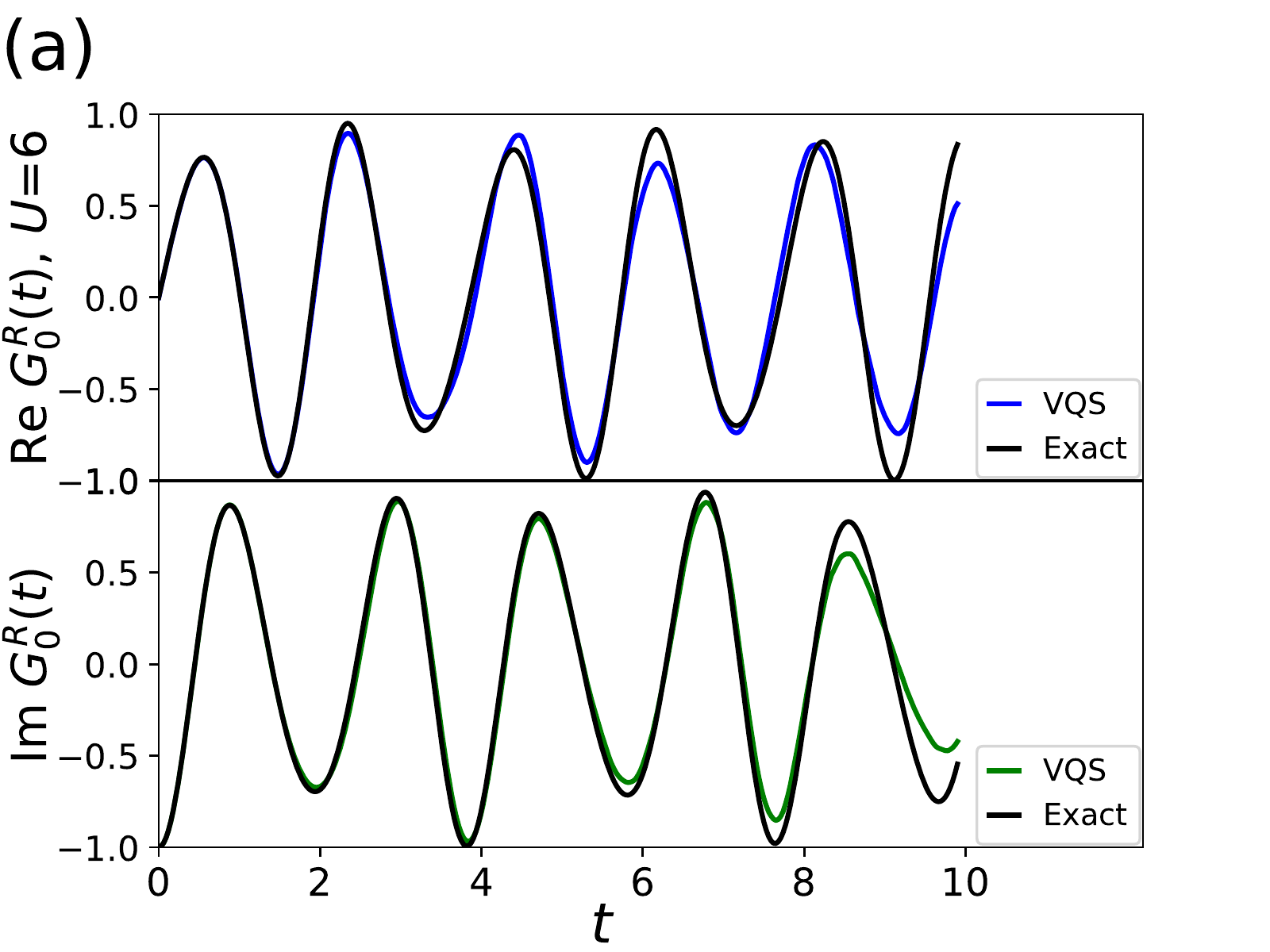}
     \includegraphics[width=.45\textwidth]{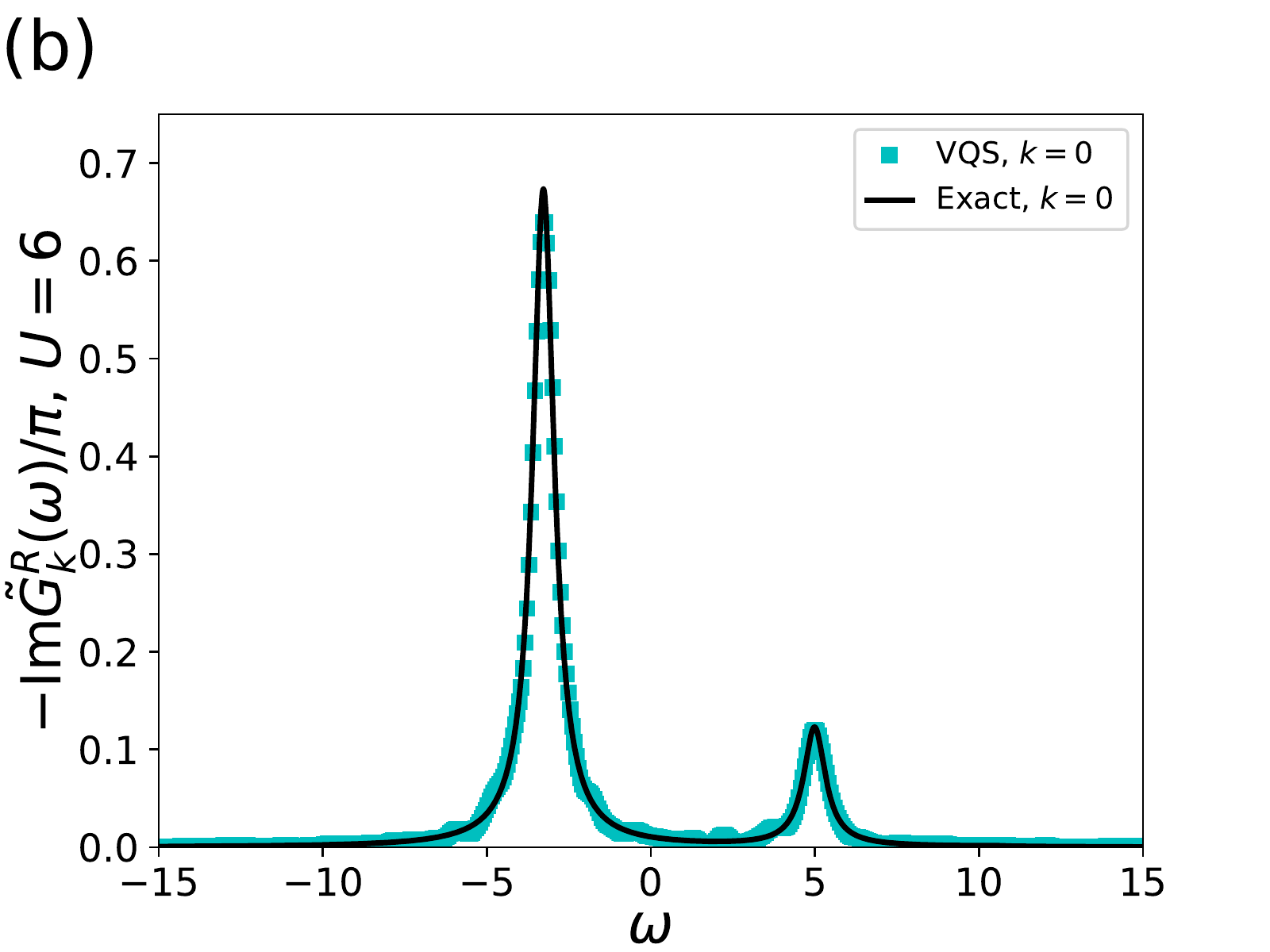}
   \caption{\label{appfig:foursitemodel}
   Numerical simulation of the VQS algorithm to compute the Green's function in real time $G^R_{k=0}(t)$ (a) and the spectral function (b) of the four-site Fermi-Hubbard model~\eqref{appeq:foursite} of $U=6$.
   The Hamiltonian ansatz~\eqref{HamAnsatz_def} of the depth $n_d=16$ is used and  
   the time step is taken as $dt=0.03$.
   The exact spectral function is calculated by the exact dynamics of the Green's function in real time from $t=0$ to $t=100$ with step $dt=0.1$.
   We take $\eta = 0.4$ for the calculation of the spectral functions.
   }
\end{figure}

Furthermore, we present results of the variational quantum simulation (VQS) for four-site Fermi-Hubbard model for further convincing the readers of the feasibility of our algorithm.
The Hamiltonian of the four-site Fermi-Hubbard model is defined as
\begin{equation} \label{appeq:foursite}
  H = - \sum_{i,=1,2,3,4,\sigma=\uparrow,\downarrow} \left( c_{i,\sigma}^\dagger c_{i+1,\sigma} + \rm{h.c.} \right)
 + U \sum_{i=1}^4 c_{i,\uparrow}^\dagger c_{i,\uparrow} c_{i,\downarrow}^\dagger c_{i,\downarrow} -\frac{U}{2} \sum_{i=1,2,3,4,\sigma=\uparrow,\downarrow} c_{i,\sigma}^\dag c_{i,\sigma}.
\end{equation}
Simulating this model requires eight qubits, so it is closer to the current experiments of the near-term quantum computers.
We employ the Hamiltonian ansatz~\eqref{HamAnsatz_def} of the depth $n_d=16$ which has 320 parameters (the number of Pauli terms in the Hamiltonian except for the identity operator is $N_P=20$).
The result of the numerical simulation of the VQS is displayed in Fig.~\ref{appfig:foursitemodel}, which well reproduces the exact result.
While general unitary operators on the system have $(2^8)^2=65536$ real parameters, our ansatz only uses real 320 parameters to simulate the dynamics for computing the Green's function by using the VQS algorithm in Sec.~\ref{sec:VQS}. 

\section{Details on numerical simulation \label{App: details} }
For the numerical calculation in Sec.~\ref{sec:VQS}, we evaluate the derivatives appearing in Eq.~\eqref{Eq:MV} by numerical differentiation by taking $\delta\theta_i=10^{-4}$ for all parameters $\theta_i$ of the ansatz quantum state.
We solve Eq.~\eqref{Eq:MthetaV} by minimizing $\| M\vec{\dot{\theta}} - \vec{V} \|$ by using \verb|numpy.linalg.lstsq| function implemented in Python library Numpy~\cite{oliphant2006guide} with neglecting the singular values of $M$ smaller than $10^{-8}$.
The size of the time step $dt$ is taken so small that there is no significant change in the final result when decreasing $dt$.

\end{document}